\documentclass[preprint,12pt]{aastex}
\begin{document}

\title{The Sloan Digital Sky Survey Quasar Catalog~V. Seventh Data Release
%\footnote{Based on observations obtained with the Sloan Digital Sky
%Survey, which is owned and operated by the Astrophysical Research Consortium.}
}
%---------Place AUTHOR.TEX after this line-------------------------------------
\author{
Donald~P.~Schneider,\altaffilmark{1}
Gordon~T.~Richards,\altaffilmark{2}
Patrick~B.~Hall,\altaffilmark{3}
Michael~A.~Strauss,\altaffilmark{4}
Scott~F.~Anderson,\altaffilmark{5}
Todd~A.~Boroson,\altaffilmark{6}
Nicholas~P.~Ross,\altaffilmark{1}
Yue~Shen,\altaffilmark{4}
W.N.~Brandt,\altaffilmark{1}
Xiaohui~Fan,\altaffilmark{7}
Naohisa~Inada,\altaffilmark{8}
Sebastian~Jester,\altaffilmark{9,10}
G.R.~Knapp,\altaffilmark{4}
Coleman~M.~Krawczyk,\altaffilmark{2}
Anirudda~R.~Thakar,\altaffilmark{11}
Daniel~E.~Vanden~Berk,\altaffilmark{1}
Wolfgang~Voges,\altaffilmark{12,13}
Brian~Yanny,\altaffilmark{14}
Donald~G.~York,\altaffilmark{15,16}
Neta~A.~Bahcall,\altaffilmark{4}
Dmitry~Bizyaev,\altaffilmark{17}
Michael~R.~Blanton,\altaffilmark{18}
Howard~Brewington,\altaffilmark{17}
J.~Brinkmann,\altaffilmark{17}
Daniel~Eisenstein,\altaffilmark{7}
Joshua~A.~Frieman,\altaffilmark{19,14,15}
Masataka~Fukugita,\altaffilmark{20}
Jim~Gray,\altaffilmark{21}
James~E.~Gunn,\altaffilmark{4}
Pascale~Hibon,\altaffilmark{22}
\v{Z}eljko~Ivezi\'{c},\altaffilmark{5}
Stephen~M.~Kent,\altaffilmark{14,15}
Richard~G.~Kron,\altaffilmark{15,14}
Myung~Gyoon~Lee,\altaffilmark{23}
Robert~H.~Lupton,\altaffilmark{4}
Elena~Malanushenko,\altaffilmark{17}
Viktor~Malanushenko,\altaffilmark{17}
Dan~Oravetz,\altaffilmark{17}
K.~Pan,\altaffilmark{17}
Jeffrey~R.~Pier,\altaffilmark{24}
Ted~N.~Price~III,\altaffilmark{4}
David~H.~Saxe,\altaffilmark{25}
David~J.~Schlegel,\altaffilmark{26}
Audry~Simmons,\altaffilmark{17}
Stephanie~A.~Snedden,\altaffilmark{17}
Mark~U.~SubbaRao,\altaffilmark{27}
Alexander~S.~Szalay,\altaffilmark{11}
and
David~H.~Weinberg\altaffilmark{28}
}

\altaffiltext{1}{
  Department of Astronomy and Astrophysics,
   The Pennsylvania State University,
   525 Davey Laboratory, University Park, PA 16802.
}
\altaffiltext{2}{
  Department of Physics,
   Drexel University, 3141 Chestnut Street, Philadelphia, PA 19104.
}
\altaffiltext{3}{
  Department of Physics \& Astronomy,
   York University, 4700 Keele Street, Toronto, Ontario, M3J 1P3, Canada.
}
\altaffiltext{4}{
  Princeton University Observatory, Peyton Hall, Princeton, NJ 08544.
}
\altaffiltext{5}{
  Department of Astronomy,
   University of Washington, Box 351580, Seattle, WA 98195.
}
\altaffiltext{6}{
  National Optical Astronomy Observatory, Tucson, AZ 85726.
}
\altaffiltext{7}{
  Steward Observatory,
   University of Arizona, 933 North Cherry Avenue, Tucson, AZ 85721.
}
\altaffiltext{8}{
  Cosmic Radiation Laboratory,
   RIKEN, 2-1 Hirosawa, Wako, Saitama 351-0198, Japan.
}
\altaffiltext{9}{
  School of Physics and Astronomy,
   University of Southampton, Southampton SO17 1BJ, UK.
}
\altaffiltext{10}{
  Max-Planck-Institut f\"{u}r Astronomie,
   K\"{o}nigstuhl 17, D-69117 Heidelberg, Germany.
}
\altaffiltext{11}{
  Department of Physics and Astronomy,
   The Johns Hopkins University,
   3400 North Charles Street, Baltimore, MD 21218-2686.
}
\altaffiltext{12}{
  Max-Planck-Institut f\"{u}r extraterrestische Physik,
   Postfach 1312, D-85741 Garching, Germany.
}
\altaffiltext{13}{
  Max Planck Digital Library, Amalienstrasse 33, D-80799, M\"unchen, Germany.
}
\altaffiltext{14}{
  Fermi National Accelerator Laboratory, P.O. Box 500, Batavia, IL 60510.
}
\altaffiltext{15}{
  Department of Astronomy and Astrophysics,
   The University of Chicago, 5640 South Ellis Avenue, Chicago, IL 60637.
}
\altaffiltext{16}{
  Enrico Fermi Institute,
   The University of Chicago, 5640 South Ellis Avenue, Chicago, IL 60637.
}
\altaffiltext{17}{
  Apache Point Observatory, P.O. Box 59, Sunspot, NM 88349.
}
\altaffiltext{18}{
  Department of Physics,
   New York University, 4 Washington Place, New York, NY 10003.
}
\altaffiltext{19}{
  Center for Cosmological Physics,
   The University of Chicago, 5640 South Ellis Avenue Chicago, IL 60637.
}
\altaffiltext{20}{
  Institute for Cosmic Ray Research,
   University of Tokyo, 5-1-5 Kashiwa, Kashiwa City, Chiba 277-8582, Japan.
}
\altaffiltext{21}{
  Microsoft Research, 301 Howard Street, No. 830, San Francisco, CA 94105.
}
\altaffiltext{22}{
  Korean Institute for Advanced Study,
   207-43 Cheongryangri-dong, Dongdaemun-gu, Seoul 130-012, Korea.
}
\altaffiltext{23}{
  Department of Physics and Astronomy,
   Seoul National University, Seoul 151-742, Korea.
}
\altaffiltext{24}{
  Division of Astronomical Sciences,
   National Science Foundation, 4201 Wilson Blvd, Arlington, VA 22230.
}
\altaffiltext{25}{
  490 Wilson's Crossing Road, Auburn, NH 03032.
}
\altaffiltext{26}{
  LBNL, Bldg 50R5032 MS 50-232, 1 Cyclotron Rd., Berkeley CA 94720-8160.
}
\altaffiltext{27}{
  University of Chicago and Adler Planetarium and Astronomy Museum,
   1300 S. Lake Shore Drive, Chicago, IL 60605.
}
\altaffiltext{28}{
  Department of Astronomy,
   Ohio State University, 140 West 18th Avenue, Columbus, OH 43210-1173.
}

%---------Place AUTHOR.TEX above this line----------------------------------

\begin{abstract}
We present the fifth edition of the Sloan Digital Sky Survey (SDSS)
Quasar Catalog, which is based upon the SDSS Seventh Data Release.
The catalog, which contains 105,783 spectroscopically confirmed
quasars, represents
the conclusion of the SDSS-I and SDSS-II quasar survey.
The catalog consists of the SDSS objects
that have luminosities larger than \hbox{$M_{i} = -22.0$} (in a cosmology with
\hbox{$H_0$ = 70 km s$^{-1}$ Mpc$^{-1}$,}
\hbox{$\Omega_M$ = 0.3,} and \hbox{$\Omega_{\Lambda}$ = 0.7),}
have at least one emission line with FWHM larger than 1000~km~s$^{-1}$
or have interesting/complex absorption features,
are fainter than \hbox{$i \approx 15.0$,} and have highly reliable redshifts.
The catalog covers an area of~$\approx$~9380~deg$^2$.
The quasar redshifts range from~0.065 to~5.46, with a median value of~1.49;
the catalog includes 1248 quasars at
redshifts greater than four, of which 56 are at redshifts greater than five.
The catalog contains 9210 quasars with \hbox{$i < 18$;} slightly over half
of the entries have \hbox{$i< 19$.}
For each object the catalog presents positions accurate
to better than~0.1$''$~rms per coordinate,
five-band ($ugriz$) CCD-based photometry with typical accuracy
of~0.03~mag, and information on the morphology and selection method.
The catalog also contains radio, near-infrared, and X-ray emission
properties of the quasars, when available, from other large-area surveys.
The calibrated digital spectra cover the wavelength region 3800--9200 \AA\ at
a spectral resolution \hbox{of $\simeq$ 2000;} the spectra can be retrieved
from the SDSS public database using the information provided in the catalog.
Over~96\% of the objects in the catalog were discovered by the SDSS.
We also include a supplemental list of an additional
207 quasars with SDSS spectra
whose archive photometric information is incomplete.

\end{abstract}


\section{Introduction}

This paper describes the Fifth Edition of the Sloan Digital Sky Survey
(SDSS; York et al.~2000) Quasar Catalog.  Previous versions of the
catalog (Schneider et al.~2002, 2003, 2005, 2007; hereafter Papers~I, II,
III, and~IV) were published
with the SDSS Early Data Release (EDR; Stoughton et al.~2002),
the SDSS First Data Release (DR1; Abazajian et al.~2003), the
SDSS Third Data Release (DR3; Abazajian et al.~2005), and the
SDSS Fifth Data Release (DR5; Adelman-McCarthy et al. 2007), and
contained 3,814, 16,713, 46,420, and 77,429 quasars, respectively.
The current catalog is the entire set of quasars from the SDSS-I and SDSS-II
Quasar Survey; the SDSS-II was completed on 15 July~2008 and the Seventh
Data Release (DR7; Abazajian et al.~2009) was made public
on 31~October~2008.  The catalog contains 105,783 quasars.

The bulk of the quasars were identified as part of the SDSS Legacy Survey,
which consisted of a systematic program to obtain spectra of
one million galaxies and 100,000 quasars in two regions: an approximately
7,600~deg$^2$ area centered on the North Galactic Pole (hereafter
this region will be referred to as the North Galactic Cap)
and approximately 800~deg$^2$ in three
narrow regions in the proximity of the Celestial Equator centered near
0$^{\circ}$ in right ascension (see York et al.~2000 and
Abazajian et al.~2009).  With this catalog, the
spectroscopy in the North Galactic Cap
region is now contiguous.  Outside of the Legacy area
quasars were found, frequently
serendipitously, on a series
of ``Special Plates" (see Adelman-McCarthy et al.~2006)
that were observed when
Legacy observations could not be performed.
A large fraction of the Special Plates
were obtained as part of two programs: Extensions of the Legacy target
selection algorithms along the Fall equatorial region
(Adelman-McCarthy et al.~2006), and
the Sloan Extension for Galactic Understanding and Exploration
(SEGUE; Yanny et al.~2009), a project to study stellar populations in the
Galactic halo.
The distribution of the DR7 quasars on the celestial
sphere is displayed in Figure~1.

The catalog in the present paper consists of quasars that
have a luminosity larger than
\hbox{$M_{i} = -22.0$}  (calculated assuming an
\hbox{$H_0$ = 70 km s$^{-1}$ Mpc$^{-1}$,} \hbox{$\Omega_M$ = 0.3,}
\hbox{$\Omega_{\Lambda}$ = 0.7} cosmology,
which will be used throughout this paper).
The objects are denoted in the catalog by their DR7
J2000 coordinates;
the format for the object name
is \hbox{SDSS Jhhmmss.ss+ddmmss.s}.  Since the image data used for the
astrometric information can change between data releases (e.g., a region
with poor seeing that is included in an early release is superseded by
a newer observation in good seeing),
the coordinates for an object can be modified at
the~$\approx$~0.1$''$ level (although on rare occasions (see \S 6) the
change in position can exceed~1$''$); hence
the designation of a given source can change between data releases.
When merging SDSS Quasar Catalogs with previous databases one should
always use the coordinates, not object names, to identify unique entries.

Recent additional compilations of quasars based upon SDSS data include
the DR5 Broad Absorption Line (BAL) Quasar Catalog (Gibson et al. 2009),
the Two Degree Field-SDSS Luminous Red Galaxy and Quasar Survey (2SLAQ)
spectroscopic catalog (Croom et al.~2009), and a sample of approximately
one million ``photometric" quasars (i.e., without spectroscopic confirmation)
selected
based upon their colors (Richards et al. 2009)
drawn from the SDSS Data Release~6 (DR6; Adelman-McCarthy et al. 2008).

The DR7 catalog does not include classes of Active Galactic Nuclei (AGN)
such as Type~2 quasars, Seyfert galaxies (i.e., low-luminosity AGN),
and BL~Lacertae objects.
Recent studies of these sources in the SDSS
can be found in Reyes et al.~(2008) (Type~2),
Hao et al.~(2005) (Seyferts), and Plotkin et al.~(2010) (BL~Lacs).
No emission line equivalent width
limit is imposed for inclusion in the present catalog;
there are some quasars with low equivalent width emission lines
in the catalog, and they tend to be among the brighter objects
(the higher the quality of the spectra, the easier the identification
of small equivalent width features), or at relatively high redshift,
where redshifts can be measured using absorption from
the Lyman~$\alpha$ forest (e.g., Fan et al.~1999).
The highest redshift SDSS quasars \hbox{($z > 5.7$;}
e.g., Fan et al.~2003, 2006; Jiang et al. 2009) were
identified as candidates in
the SDSS imaging data, but the spectra were not obtained with the
SDSS spectrographs,
so they are not included in the catalog.

The basic format and analysis of this work closely follow that
of the previous four editions of the SDSS Quasar Catalog.
The observations used to produce the catalog are presented in
Section 2; the construction of the catalog and the catalog format
are discussed in Sections 3 and~4, respectively.  Section~5
presents an overview of the catalog, Section~6 contains
a discussion of some issues relevant
to statistical samples, and a summary is given in Section~7.
The catalog is presented in an electronic table in this paper and
can also be found at an SDSS public web
site.\footnote{\tt
http://www.sdss.org/dr7/products/value$\_$added/qsocat$\_$dr7.html}

\section{Observations}

The Sloan Digital Sky Survey
used a CCD camera \hbox{(Gunn et al. 1998)} on a
dedicated \hbox{2.5-m} telescope
\hbox{(Gunn et al. 2006)}
at Apache Point Observatory,
New Mexico, to obtain images in five broad optical bands ($ugriz$;
Fukugita et al.~1996) over approximately
10,000~deg$^2$ of the high Galactic latitude sky.
The
survey data-processing software measures the properties of each detected object
in the imaging data in all five bands, and determines and applies both
astrometric and photometric
calibrations (Pier et al., 2003; Hogg et al.~2001; \hbox{Lupton et al. 2001};
Ivezi\'c et al.~2004).  The details of the SDSS photometric calibration
are described in Smith et al.~(2002), Tucker et al.~(2006), and
Padmanabhan et al.~(2008).
The SDSS photometric system is based on the AB magnitude scale
(Oke \& Gunn~1983); the photometric measurements are reported as
asinh magnitudes (Lupton, Gunn, \& Szalay~1999).
Unlike previous editions of the SDSS quasar
catalog, in this release we use the ``ubercalibration" photometric
calibrations of Padmanabhan et al.~(2008).

The SDSS quasar target selection is described in Richards et al.~(2002,
hereafter GTR02) and
in papers I-IV; we provide a brief summary of the identification of candidates
and the spectroscopic observations here.
Most quasar candidates are selected based on
their location in multidimensional SDSS color-space.
The Point Spread Function (PSF) magnitudes are used for the quasar
target selection, and the selection is based upon measurements
that have been corrected for Galactic extinction
using the maps of Schlegel, Finkbeiner, \& Davis~(1998).
An $i$ magnitude limit of~19.1
is imposed for candidates whose colors indicate
a probable redshift of less than~$\approx$~3 (selected from the $ugri$
color cube);
high-redshift candidates (mainly selected from the $griz$ color cube,
but some $z<3.5$ are identified in the $ugri$ color cube; see GTR02)
are accepted if \hbox{$i < 20.2$} and the source is unresolved.
In addition to the multicolor selection, unresolved objects brighter
\hbox{than $i = 19.1$} that lie within~2.0$''$ of a FIRST radio source
(Becker, White, \&~Helfand~1995) are also identified as primary quasar
candidates.
Target selection also imposes a maximum brightness limit
\hbox{($i \approx 15.0$)} on quasar candidates to avoid saturation and
cross-talk in the spectra.

The primary sample described above was supplemented by
quasars that were targeted by
the following SDSS spectroscopic target selection algorithms:
Galaxy and Luminous Red Galaxy (Strauss et al.~2002 and
Eisenstein et al.~2001),
X-ray (object near the position of a {\it ROSAT} All-Sky Survey
[RASS; Voges et al.~1999,~2000] source; see Anderson et al.~2003),
Star (point source with a color typical of an interesting class of star),
or Serendipity (unusual color or FIRST matches).  The SDSS is designed to
be complete in the Galaxy, Luminous Red Galaxy, and Quasar programs,
but no attempt at completeness was made for the other three categories.
Figure~2 shows the locations of stars and quasars in the SDSS color-color
space (the quasar measurements are from objects in the DR7 catalog).

The final quasar selection algorithm of GTR02 was refined using
the results of early SDSS observations, so not all objects in this
catalog were selected using the GTR02 approach; all quasars originally
reported in Papers~I and~II, and some in Paper III, were identified with
pre-GTR02 methods.
%The selection for the UV-excess ($z < 2$) quasars,
%which comprise~$\approx$~80\% of the
%DR7 Catalog, has remained
%reasonably uniform; the changes to the selection algorithm were primarily
%designed to increase the effectiveness of the identification of
%\hbox{$3.0 < z < 3.8$} quasars.
The catalog reports two target selection
flags for each object: 1)~TARGET, which is the flag that was used, based
upon the imaging information and the selection algorithm available at the
time, to
select the quasar for the spectroscopic sample, and 2)~BEST, which is the
flag produced with the GTR02 algorithm and the latest available
photometric information.
%Note that the TARGET and BEST values for the
%target selection flag may differ
%even though the quasar was selected with the GTR02 algorithm; if the
%photometric values change either because of new imaging data or a reprocessing
%of the imaging data with a new version of the photometric or calibration
%pipeline, the BEST target flag can change.
%The possibility of updated photometry for an object gives rise to two sets
%of photometric measurements for each quasar: one set
%used at the time of the spectroscopic target selection, the other set
%refers to the measurements with the latest photometric software
%on the highest quality data.  The first group is designated as TARGET
%photometry, the second as BEST photometry.
%The TARGET selection flag is based on the TARGET photometry, the
%BEST target values are based upon the BEST photometric measurements.
Care must be exercised when constructing statistical samples
from this catalog;
%if one uses the BEST values
%of the selection software, not only must one drop the catalog
%quasars that were not targeted
%as quasar candidates by BEST, one must also somehow
%account for
%quasar candidates selected by the final version that were not observed in the
%SDSS spectroscopic survey.  To create a statistical sample, one should always
%use the TARGET values;
see GTR02, Vanden~Berk et al.~(2005),
and Richards et al.~(2006) for discussions of survey completeness and
efficiency, and the issues that are
important for the construction of
statistical SDSS quasar samples.

In addition to the quasars found in the Legacy Survey, the catalog contains
many quasars whose spectra were obtained on Special Plates
(see Adelman-McCarthy et al.~2006).  The main source
of Special Plate quasars is the set of observations along the celestial
equator in the Southern Galactic Cap (``Stripe~82"; see
Stoughton et al.~2002) taken for a variety of commissioning and science
purposes; for example, exploring the variations in the selection criteria
of spectroscopic targets and empirically determining the completeness of the
quasar survey (e.g., Vanden Berk et al.~2005).
Another major component of the Special Plates are
the observations taken for the SEGUE project (Yanny et al.~2009); these
plates usually lie along lines of constant Galactic longitude and reach
closer to the Galactic plane than do the Legacy Survey fields (see Figure~1).
The SEGUE program required SDSS imaging outside of the Legacy Survey area.

The spectroscopic targets identified by the various SDSS selection algorithms
are arranged onto
a series of 3$^{\circ}$ diameter circular fields (Blanton et al.~2003).
The two SDSS double spectrographs produce data covering \hbox{3800--9200 \AA }
at a spectral resolution \hbox{of $\simeq$ 2000;} a dichroic splits the beam
at~6150~\AA.
The data, along with the associated calibration frames, are processed by
the SDSS spectroscopic pipeline (see Stoughton et al.~2002).  Several
improvements have been made to the spectroscopic software since DR5, in
particular the wavelength calibration, spectrophotometric calibration, and
the handling of strong, unresolved emission features (see
Adelman-McCarthy et al.~2008 and Abazajian et al~2009
for a description of these modifications).

The calibrated spectra are classified into various groups
(e.g., star, galaxy, quasar), and redshifts are determined by two independent
software packages: {\tt spectro1d} (see \S 4.10.2.1 of Stoughton et al.~2002),
which assigns absorption-line redshifts by cross-correlating the
observations with a series of template spectra in Fourier space and
Gaussian fits to emission lines to measure emission-line redshifts; and
{\tt specBS} (Adelman-McCarthy et al.~2008), which performs $\chi^2$ fits
in wavelength space between the observations and templates.
The quality of the redshift is quantified by {\tt spectro1d} by a parameter
called the ``Confidence Limit", which is stored in the parameter {\tt zconf}
(Stoughton et al.~2002).
A large value of {\tt zconf} usually indicates a solid redshift measurement,
but a low value does not necessarily mean an uncertain redshift; see
\S 6 for a discussion of this important issue.
The catalog reports the {\tt spectro1d} redshifts,
which have typical quoted redshift errors of~$\approx$~0.004 (except in the
cases where visual inspection found that the {\tt spectro1d} redshift was
incorrect; see \S 3.2).
%The redshifts produced by {\tt spectro1d} and {\tt specBS} agree to
%3000~km~s$^{-1}$ for~$\approx$~91\% of the bona-fide quasars that
%both software packages
%classify as quasars.
%A comparison of 299 quasars observed
%at multiple epochs by the SDSS (Wilhite et al.~2005) found an rms
%difference of~0.006 in the {\tt spectro1d} redshifts for a given object,
%consistent with the quoted errors.
Hewett \& Wild (2010) present evidence for systematic errors of~$\approx$~0.003
in SDSS quasar redshifts measurements; these errors are small for most
applications, but can be significant for some investigations.

The catalog contains photometry from~346
SDSS imaging runs acquired between
19~September~1998 (Run~94) and 1~January~2008 (Run~7264),
and spectra from 2210 spectroscopic plates taken between
5~March~2000 and 6~July~2008.
Figure~3 shows the calibrated SDSS spectra of four
catalog quasars, added since Paper~IV, representing a range of properties.
The processed DR7 spectra {\it have not} been corrected for Galactic
extinction.

\section{Construction of the SDSS DR7 Quasar Catalog}

As described above, the quasars in the catalog were drawn from two
sets of SDSS observations: the Legacy Survey area,
and the Special Plates, where the spectroscopic targets were not chosen
by the standard SDSS target selection algorithms.
The DR7 quasar catalog was constructed in three stages: 1)~creation of a
quasar candidate database, 2)~visual examination of the spectra of the
quasar candidates, and 3)~application of luminosity
and emission-line velocity width criteria.  This approach is similar to that
used in Papers I-IV, but with minor modifications.

%\subsection{Creation of the DR7 Quasar Candidate Database}
%
This catalog of bona-fide quasars, that have redshifts checked by eye
and luminosities and line widths that meet the formal quasar
definition, is constructed from a larger ``master" table of quasar
candidates and confirmed quasars.  This master table was created using an
SQL query to the public SDSS-DR7 database (i.e., the Catalog Archive
Server [CAS]; see Thakar et al 2008 and {\tt http://cas.sdss.org/astrodr7/}).
%Two versions of the
%photometric database exist, which contain the properties of objects
%when targeted for spectroscopic observations (TARGET) and as
%determined in the latest processing (BEST).  These databases are
%divided into multiple tables and subtables to facilitate access to
%the most relevant data for a particular use.  In the case of the
%quasar catalog construction, we have made use of the {\tt PhotoObjAll}
%and {\tt SpecObjAll} tables, which contain, respectively, the
%photometric information for {\em all} SDSS sources detected in the images
%and for {\em all}
%SDSS spectra.  In the case of {\tt PhotoObjAll}, both the TARGET and
%BEST versions are queried.  These tables are the most complete database
%files, and include duplicate
%observations of objects and observations of objects that lie outside
%of the Legacy footprint (as compared to the {\tt PhotoObj} and {\tt SpecObj}
%tables, which include only sources in the Legacy area).
%In {\tt PhotoObjAll}, two (or more)
%observations of a single object may exist;
%if so, one is classified as {\tt PRIMARY}, the other(s) as {\tt SECONDARY}.
%
The vast majority of the objects in the master table were found from the
union of four
classes of objects: 1)~those targeted, by either TARGET or BEST,
by the quasar selection
algorithm (GTR02) based upon their photometric or radio properties,
2)~those whose spectra were classified by {\tt spectro1d} as quasars
({\tt specClass=QSO} or {\tt HIZ\_QSO}),
3)~those whose spectra are classified as {\tt UNKNOWN} by {\tt spectro1d},
and 4)~any object whose {\tt spectro1d} redshift is larger
than~0.6. (There is of course considerable
overlap between these categories.)  The spectra in the third category
often have low S/N or some glaring instrument-induced defects, but the
set of {\tt UNKNOWN} objects also
include a considerable number of quasars that
would have been missed by the other three selection criteria (e.g., extreme
BALs; Hall et al.~2002).  The final
category recovered the few quasars that were not selected by the GTR02
algorithm but whose spectra were misclassified by
{\tt spectro1d}, usually as
galaxies.  The lower redshift limit of criterion (4)
was imposed because of the rapid rise in the number of objects
as the redshift limit was decreased.  A search of the {\tt specBS} database
found 112 quasars that had not been identified by the SQL query; these objects
were added to the master quasar list.

%Quasar candidates are those objects which have had one or
%more of the following flags set by the algorithm described by GTR02
%
%\noindent {\tt TARGET\_QSO\_HIZ OR TARGET\_QSO\_CAP OR
%TARGET\_QSO\_SKIRT OR TARGET\_QSO\_FIRST\_CAP\\ OR
%TARGET\_QSO\_FIRST\_SKIRT}
%
%\noindent \hbox{( = 0x0000001F,} except for the Special Plates,
%where additional care is required in interpreting the CAS flags).
%Objects flagged as {\tt TARGET\_QSO\_MAG\_OUTLIER} and {\tt
%TARGET\_QSO\_REJECT} are not included, as these flags are meant only
%for diagnostic purposes; the former sources fall outside of the magnitude
%limits of the quasar survey, but are viewed as promising quasar candidates,
%whereas the latter term refers to those objects that
%lie in locations of color-color
%space where the contamination from non-quasars is expected to be high.
%(These flags are described in the EDR paper;
%in the CAS documentation and the EDR paper,
%{\tt TARGET\_QSO\_MAG\_OUTLIER} is called {\tt TARGET\_QSO\_FAINT}.)

Comparison of a preliminary master table to the 77,429 quasars included in
Paper~IV revealed
that a small number of quasars were missing, mostly due to changes in adopted
redshifts or classifications.  The following subsets were added to the master
table, using purely heuristic criteria designed to maximize the recovery of
known missing objects while minimizing the inclusion of large numbers of
objects that were not quasars.

\noindent
$\bullet$~Objects whose cross-correlation with quasar templates in the
{\tt spectro1d} processing yielded \hbox{$z > 0.6$,} but for which the
CAS entry is a higher-confidence {\tt spectro1d} redshift below~0.6
based upon fitting to emission lines,
and with a confidence on that cross-correlation redshift which
is at least~80\% of the CAS {\tt zConf} value.

\noindent
$\bullet$~Objects whose {\tt spectro1d} emission-line redshift exceeded~0.6,
but for which the
CAS entry is a higher-confidence {\tt spectro1d} redshift below~0.6
based upon cross-correlation with the quasar templates,
and with a confidence on that emission-line cross-correlation redshift which
is at least~80\% of the CAS {\tt zConf} value.

\noindent
$\bullet$~Objects at $0.35<z<0.6$ which have a \ion{Mg}{2}
emission line detected at $\geq 3.5\sigma$ significance with an observed
linewidth (FWHM)
of $\geq$ 5.4\,\AA\ and a linewidth uncertainty of no more than 40\%.

\noindent
$\bullet$~One Paper~IV quasar is not
in the DR7 CAS (the ``unmapped fiber" object discussed in Paper~IV);
information on this
object, taken from Paper~IV, was manually added to the master table.

The final table contains information for 1,293,260
sources\footnote{The
master table is known as the
QSOConcordanceAll table, which can be found in the SDSS database; see
$\mbox{\tt
http://cas.sdss.org/astrodr7/en/help/browser/description.asp?n=QsoConcordanceAll\&t=U.}$},
of which 300,357 entries have spectroscopic observations.
Multiple entries for a given object are retained in the master table.
The following culls were then applied to this database:
1)~Objects that were lacking spectra (992,903 entries),
2)~Objects whose photometric measurements have not been loaded into the CAS,
and thus luminosity calculations are impossible to perform (6383;
207 of these objects are indeed quasars and are described in Appendix~A),
3)~Candidates that were classified by {\tt spectro1d} as
{\tt SPEC\_STAR} or {\tt SPEC\_SKY}, with redshifts less than~0.002, and
had the redshift status flag set to either {\tt XCORR\_EMLINE},
{\tt XCORR\_HIC},
{\tt EMLINE\_XCORR}, or {\tt EMLINE\_HIC} (48,712),
and 4)~Multiple spectra (coordinate agreement better than~1.0$''$)
of the same object~(65,632).

Before proceeding, we should make a few brief comments on the last two
items.  Ten of the Paper~IV quasars were classified by the DR7 spectroscopic
pipeline as stars with high confidence.  In these individual cases, we
overrode the automatic rejection and retained the quasars in the catalog.
There are undoubtedly a few additional quasars residing in this large subset
of confidently classified stars, but the effort required to visually examine
the nearly 50,000 spectra was not deemed to be
worth the miniscule return of quasars.
In cases of duplicate spectra of an object,
the ``science primary" spectrum is selected (i.e., the spectrum was obtained as
part of normal science operations);
when there is more than one science primary observation (or when none of
the spectra have this flag set), the spectrum with the
highest S/N is retained (see Stoughton et al.~2002
for a description of the
science primary flag).  These actions produced a list of~179,630
unique quasar candidates.

%\subsection{Visual Examination of the Spectra}

Given the considerable size of the DR7 quasar candidate list, our previous
practice of visually examining every spectrum, irrespective of whether
it had been reviewed for a previous edition of the catalog, was abandoned.
Only those objects whose DR7 spectra (identified by the Modified Julian
Date of observation, plate number, and fiber number) were not in the
Paper~IV list were by default reviewed (55,666 spectra).  We also visually
examined the spectra for the 1,224 quasars whose Paper~IV redshifts differed
by the DR7 {\tt spectro1d} values by more than~0.03.

The SDSS spectra of these quasar candidates were manually inspected by
several of the authors;
%(DPS, PBH, GTR, MAS, SFA, YS, NI, and PH);
as in
previous papers in this series, we found that the
spectroscopic pipeline redshifts and classifications
of the overwhelming majority of the objects
are accurate.
The redshifts for 2671 of the quasar candidates
were manually adjusted, usually by significant (several tenths or more)
amounts; the main cause for revision, as one would expect, was that
{\tt spectro1d} assigned the incorrect identification to emission features.
%Tens of thousands of objects were dropped from the
%list because they were obviously not quasars (e.g., no emission lines and
%not BALs; no broad emission lines) or their spectra lacked
%sufficient S/N to assign a plausible redshift.
%This large number of rejections
%is not unexpected given the exceedingly
%wide net cast in our construction of the master table.

%The catalog contains numerous examples of
%extreme Broad Absorption Line (BAL) Quasars
%(see Hall et al.~2002); the spectrum of one of the objects in this
%class can be seen in Figure~3.  Although they occasionally fail the
%quasar emission-line width criterion, BALs have
%historically been included in quasar
%catalogs, so we have retained objects in this class that
%meet the \hbox{$M_i < -22.0$}
%luminosity criterion.

%\subsection{Luminosity and Line Width Criteria}

The absolute magnitude limit for inclusion in the catalog is
\hbox{$M_{i} = -22.0$,}
calculated by correcting the BEST $i$
PSF magnitude measurement for Galactic extinction (using the maps of
Schlegel, Finkbeiner, \& Davis~1998) and assuming that the quasar
spectral energy distribution in the ultraviolet-optical
can be represented by a power law
\hbox{($f_{\nu} \propto \nu^{\alpha}$),} where $\alpha$~=~$-0.5$
(Vanden~Berk et al.~2001).  In the 20 cases where BEST $i$ photometry
was unavailable, the TARGET measurements were substituted for the
absolute magnitude calculation.  (There are 52 quasars that lack TARGET
photometry.)
%An object of $M_i = -22.0$ will reach
%the \hbox{$i = 19.1$} ``low-redshift" ($ugri$)
%selection limit at a redshift of~$\approx$~0.35 and
%the \hbox{$i = 20.2$} ``high-redshift" ($griz$) cutoff
%at a redshift of~$\approx$~0.54.  This latter redshift is far below
%the redshift targeted by the $griz$ selection process.

This approach ignores the contributions
of emission lines
and the observed distribution in continuum slopes to the photometric
measurements.
Emission lines can contribute several tenths of a magnitude to the
k-correction (see Richards et al. 2006), but since there is essentially
no emission line contribution to the $i$ flux at zero redshift, our
adopted approach will be inclusive (i.e., if one wishes to apply a
k-correction
for emission lines to construct a sample of objects with
continuum
luminosities greater than or equal to \hbox{$M_{i} = -22.0$,} our luminosity
cut will not have removed any objects that satisfy the
emission-line k-corrected sample).
%The calculated
%absolute magnitudes may also be considerably in error because of the
%variations in the continuum slopes (this error also depends upon the
%redshift of the object).
%The absolute magnitudes
%will be particularly uncertain (a magnitude or more)
%at redshifts near and above
%five, when the Lyman~$\alpha$ emission line (with a typical observed
%equivalent width \hbox{of $\approx 400-500$ \AA })
%and strong Lyman~$\alpha$ forest absorption enter
%the~$i$ bandpass.  At these redshifts, however, all detected quasars must have
%true luminosities far above \hbox{$M_{i} = -22.0$.}

Quasars near the $M_i$~=~$-22.0$ luminosity
limit are often of similar brightness in the $i$-band as the
starlight produced by the host galaxy.  Although the PSF-based
SDSS photometric data presented
in the catalog are less susceptible to host galaxy contamination than
are fixed-aperture measurements, the nucleus of the host
galaxy can still
contribute appreciably to this measurement for the lowest luminosity
entries in the catalog (e.g., Hao et al.~2005, Bentz et al.~2006,
Croom et al.~2009).

The visual inspection and application of the luminosity criterion
reduced the number of quasar candidates to under 106,000 objects.  During the
visual inspection we included all spectra that might possess at least one
broad
line, but no quantitative linewidth measurements were made.
To remove narrow-line
objects, the spectra of the candidate quasars were
examined with
the Principal Component Analysis (PCA) code of Boroson \& Lauer~(2010).
The PCA analysis was done by constructing eigenspectra from
approximately 1000 of the highest S/N spectra after rejecting the few
narrow-lined objects in this subset.

This procedure was applied separately to the 4300--5400~\AA\ rest frame
region of
the 15,900 objects with $z<0.7$ and to the 2400--3200~\AA\ rest frame
region of the 55,400 objects with \hbox{$0.7<z<1.8$.}
The spectra that
could not be fit with a small number (15) of these eigenspectra were
visually examined.  Almost all of the 120 rejects were narrow-line objects,
with about 75\% being Seyfert~2s or LINERS, and the remainder were spectra
with H~II-region line ratios.  We also examined the spectra of sources
that were located near the highest concentration of narrow-line objects
in the eigenspace; this exercise identified a handful of additional objects
to reject.  This process also uncovered a few objects with incorrect
redshifts, which we revised.

Several of the high-redshift ($z>0.7$) quasar candidates have images
that appear extended, as quantified by the difference between their
PSF and model magnitudes (see, e.g., the discussion in Stoughton et
al. 2002; Strauss et al. 2002).  We examined in detail the spectra and
images of all $z>0.7$ quasar candidates with $r$ band PSF - model magnitudes
greater than~0.2 mag.  Many of these objects are
quasar-galaxy superpositions, and the sample includes previously known
gravitational lenses (e.g., Inada et al. 2010; McGreer et al. 2010).
There were also 24 objects which on closer inspection were not quasars,
and they were removed from the sample.

\section{Catalog Format}

The DR7 SDSS Quasar Catalog is available in three types of files at the
SDSS public web site listed in the introduction:
1)~a standard ASCII file with fixed-size columns,
2)~a gzipped compressed version of the ASCII file,
and 3)~a binary FITS table format.
The following description applies to the standard ASCII file.  All files
contain the same number of columns, but the storage of the numbers differs
slightly in the ASCII and FITS formats; the FITS header contains all of the
required documentation.

The standard ASCII catalog (Table~2 of this paper), which is~52~MB in size,
has a format that differs in only minor ways from that used in Paper~IV.
The first~80 lines consist of catalog documentation; this material is followed
by~105,783 lines, each containing~74 columns.  A summary
of the information is given in Table~1 (the documentation in the ASCII catalog
header
is essentially an expansion of Table~1).  At least one space separates all the
column entries, and, except for the first and last columns (SDSS designation
and the object name if previously known),
all entries are reported in either floating point or integer format.

A summary of the spectroscopic selection, for both the TARGET and the BEST
algorithms, is given in Table~3.  We report
seven selection classes in the catalog (columns~38 to~44 for BEST,
\hbox{55--61} for TARGET).  Each selection version has two columns,
the number of objects that
satisfied a given selection criterion and
the number of objects that were identified only by that selection
class.
Of the~70,365 DR7 quasars that have Galactic-absorption corrected TARGET
$i$ magnitudes brighter than~19.1,
68,497 (97.3\%) were identified by the TARGET quasar multicolor
selection; if one combines TARGET multicolor and FIRST selection (the primary
quasar target selection criteria),
all but~1150 of the \hbox{$i < 19.1$} objects were selected.  (The spectra
of many of the last category of objects were obtained in observations that
were not part of the primary quasar survey, i.e., the Special Plates.)
The numbers
are similar if one uses the BEST photometry and selection, although the
completeness is not quite as high as with TARGET values.

\section{Catalog Summary}

The 105,783 objects in the catalog represent an increase of 28,354 quasars
over the Paper~IV database; of the entries in the new catalog,
101,945 (96.4\%) were either not listed in the NASA/IPAC Extragalactic
Database (NED) or were recorded as an SDSS
discovery in the NED database. (NED occasionally
lists the SDSS name for objects that were not discovered by the SDSS.)
The catalog quasars span a wide range of properties: redshifts
from~0.065 to~5.461, \hbox{$ 14.86 < i < 22.36$}
(1341~objects \hbox{have $i > 20.5$;} only~183
have \hbox{$i > 21.0$}),
and \hbox{$ -30.28 < M_{i} < -22.00$.}
The catalog contains 8630, 5377, and~53,564
matches to the FIRST, RASS, and 2MASS data, respectively.
The RASS and 2MASS catalogs cover essentially all of the DR7 area, but~6865
(6.5\%) of the entries in the catalog lie outside of the FIRST region.

Figure~4 displays the distribution of the DR7 quasars in the
$i$-redshift plane.
Objects denoted by NED as previously discovered by investigations other
than the SDSS
are indicated with open circles.  The curved cutoff on the left
hand side of the graph is produced by the minimum luminosity criterion
\hbox{($M_i < -22.0$).}  The ridge in the contours at
\hbox{$i \approx 19.1$}
for redshifts below three reflects the flux limit of the
low-redshift sample; essentially all of the
large number of \hbox{$z < 3$} points with \hbox{$i > 19.1$}
are quasars selected via criteria other than the primary
multicolor sample.

The clear majority of
quasars have redshifts below two (the median redshift is~1.49, the
mode is~$\approx$~1.85),
but there is a significant tail
of objects extending to redshifts well beyond $z>4$ (see upper left panel
in Figure~5).
%The dips in the curve at redshifts
%of~2.7 and~3.5 arise because the SDSS colors of quasars at these redshifts
%are similar to the colors of stars; we decided to accept significant
%incompleteness at these redshifts rather than be overwhelmed by a large number
%of stellar contaminants in the spectroscopic survey.  Improvements in the
%quasar target selection algorithm since the initial editions of the
%SDSS Quasar Catalog have increased the efficiency of target
%selection at redshifts near~3.5 (compare this panel with Paper~II's
%Figure~6; see GTR02 for a discussion of the incompleteness
%of the SDSS Quasar Survey).
%
The catalog contains~152 quasars with redshifts below~0.15
(21 \hbox{with $z < 0.10$}).
All of the 152 objects are
of low luminosity \hbox{($M_i > -24.0$, only six have $M_i < -23.5$)}
because of the \hbox{$i \approx 15.0$}
limit for the spectroscopic sample.  About 70\%
of these quasars (106) are classified by the SDSS processing software
as resolved in the SDSS image data.
The images of~4392 of the quasars are classified as extended by the
SDSS photometric pipeline;
4124~(94\%) have redshifts below one
(there are ten resolved \hbox{$z > 3.0$} quasars).
%A total of~48
%of the \hbox{$z < 0.15$} quasars were discovered by the SDSS.
The spectrum of the lowest redshift object is displayed
in Figure~3.
The DR7 catalog contains 1248 quasars with redshifts larger than four;
56 entries have redshifts above five, and one of the post-DR5 quasars has
a redshift of~5.461 (see Figure~3), which exceeds the previous record of~5.41
found in the SDSS spectroscopic survey (Anderson et al.~2001).
%At redshifts larger than~$\approx$~5.7
%the observed wavelength of the
%Lyman~$\alpha$ emission line is redward of
%the $i$ band; at this point quasars become single-filter ($z$) detections
%in the SDSS imaging survey.
%At the typical $z$-band flux levels for redshift six quasars, there are
%too many ``false-positives" to obtain spectra as part of routine SDSS
%operations.
%
%\subsection{Observed Magnitude and Luminosity Distributions}
%
%The distribution of the observed $i$ magnitude
%(not corrected for Galactic extinction) of the quasars shows a sharp
%drop (much more pronounced in a linear scaling) at
%\hbox{$i \approx 19.1$;} this arises because of the magnitude limit for the
%low-redshift sample (GTR02).  A second discontinuity, at
%\hbox{$i \approx 20.2$,} is due to the high-redshift sample magnitude limit;
%quasars fainter than the
%\hbox{$i = 20.2$} high-redshift were identified via other
%selection algorithms, primarily serendipity.

The $i$-magnitude distribution of the DR7 quasars is shown in the upper right
panel of Figure~5.
Although the spectroscopic survey is limited to objects fainter than
\hbox{$i \approx 15$}, the SDSS discovered a number of
``PG-class"
(Schmidt \& Green~1983) objects (see Jester et al.~2005).
The DR7 catalog contains 124 entries
with \hbox{$i < 16.0$}; 19 of the quasars lack a NED entry.
As can be seen from the $M_i$ diagram (lower right panel in Figure~5),
the SDSS quasar survey is composed
primarily of fairly luminous quasars; the distribution is roughly symmetric
around \hbox{$M_i = -26$} with a FWHM
of approximately one magnitude
(in the adopted cosmology
3C~273 has \hbox{$M_i \approx -26.6$).}
The histogram declines sharply at
high luminosities (only~1.6\% of the objects have \hbox{$M_i < -28.0$)}
and has a gradual decline toward lower luminosities.
Of the 134 catalog quasars with \hbox{$M_i < -29.0$,}
84 either have no NED entry or are identified as an SDSS quasar;
the redshifts of these objects lie
between~1.30 and~4.97.  Six quasars have \hbox{$M_i < -30.0$.}
The two most luminous quasars have virtually
identical absolute magnitudes ($-30.26$ and~$-30.28$):
\hbox{2MASSI J0745217+473436} \hbox{(= SDSS J074521.78+473436.2),}
\hbox{at $z = 3.22$,} and \hbox{HS 1700+6416}
\hbox{(= SDSS J170100.60+641209.3),}
\hbox{at $z = 2.735$.}

%The majority of the large-redshift ``resolved"
%quasars are probably measurement
%errors, but this sample may also contain a mix of chance superpositions
%of quasars and foreground objects and some
%small angle separation gravitational lenses (indeed, several lenses
%are present in the resolved quasar sample; see Oguri et al.~2006 and
%Inada et al.~2008).

%
%\subsection{Broad Absorption Line Quasars}

%The SDSS quasar selection algorithm has proven to be effective
%at finding a wide variety of Broad Absorption Line (BAL) Quasars;
%previous compliations of SDSS BAL quasars are
%Tolea, Krolik, \& Tsvetanov~(2002),
%Reichard et al.~(2003), Trump et al.~(2006) and Gibson et al.~(2009).
%The final reference presents the BAL properties of~5039 quasars included
%in Paper~IV.
%BAL quasars are usually recognized by the presence of
%C~IV absorption features, which are only visible in SDSS spectra at
%$1.6 < z < 4.9$.  (The much rarer Mg~II BALs can be identified at
%$0.4 < z < 2.2$ in SDSS spectra.)
%We expect that there are approximately 2000 BAL quasars
%in the catalog that have been added since Paper~IV.
%The SDSS has discovered a wide variety of extreme BAL quasars
%(see Hall et al.~2002); see the lower left panel in Figure~3 for an
%example of this type of object.

It has long been known that the majority of optically-selected
quasars inhabit a restricted range
in photometric color, and the large sample size and
accurate photometry of the SDSS revealed a tight color-redshift
correlation for quasars (Richards et al.~2001, 2003, Paper~IV).
%The dependence of the four standard SDSS colors on redshift for the
%DR7 quasars is given in Figure~6.  ( All photometric measurements used
%in these analyses have been corrected for Galactic extinction.)
%The solid line in each panel is the
%modal relation for the DR7 quasars; the modal relations are tabulated in
%Table~5, along with the values for~$(g-i)$ (see Paper~IV for details of the
%calculation).  The relations are not defined for redshifts below~0.12
%or above~5.12 because of the low numbers of quasars in this range.
%All four panels show an impressively
%tight correlation of color with redshift; the increased scatter at
%high redshifts occurs when
%the Lyman~$\alpha$ forest begins to dominate the bluer of the passbands
%used to form the color.  The color-redshift relations in Table~5
%are not significantly different from the values reported in Paper~IV,
%but the number of quasars used to create the relation has
%increased considerably.
%
The modal relations for quasar
colors in the four standard SDSS colors as a function of redshift are
given in Paper~IV; the differences between those values and the ones
calculated from the DR7 sample are negligible.
For each of the catalog quasars we provide the quantity $\Delta (g-i)$, which
is defined by

$$ \Delta (g-i) \ =
\ (g-i)_{\rm QSO} \ - \ \langle (g-i) \rangle_{\rm redshift} $$

\noindent
where $\langle(g-i) \rangle_{\rm redshift}$ is the modal value of this color
at the redshift of
the quasar.  This ``differential color" provides an estimate of the relative
continuum slope
of the quasar (values above zero indicate that the object has
a redder continuum than the typical quasar at that redshift).
The distribution of $\Delta (g-i)$ near the modal curve is roughly
symmetric, but the lower left panel in Figure~5 reveals a significant
population of ``red" quasars
that has no ``blue" counterpart (also see Richards et al.~2003 and
York et al.~2006).
The usefulness of this measurement is compromised at redshifts above three
because of the impact of the Lyman~$\alpha$ forest and Lyman-limit systems
in the observed $g$ photometry.
These color-redshift relations
have led to considerable
success in assigning photometric redshifts to quasars (e.g., Weinstein
et al.~2004 and references therein).
%
%
%\subsection{Matches with Non-optical Catalogs}

A total of 8630 catalog objects are FIRST (Becker, White, and Helfand~1995)
sources (defined by a SDSS-FIRST
positional offset of less than~2.0$''$).  Note that 336 of the objects
were selected (with TARGET) solely because
they were FIRST matches.
Extended radio sources may be missed by this matching.  The upper left
panel in Figure~6 contains a histogram of the angular offsets between the
SDSS and FIRST positions; the solid line is the expected distribution assuming
a~0.20$''$ $1\sigma$ Gaussian error in
the relative SDSS/FIRST positions (found by fitting only the points with a
separation less than~1.0$''$).  The small-angle separations are well-fit to
the Rayleigh distribution,
but outside of about~0.5$''$ there is an obvious
excess of observed separations; this large-separation tail is presumably due
to sources with extended and/or complex radio emission.
Given the high astrometric accuracy of the
SDSS and FIRST catalogs, the fraction of false matches is quite small
(on the order of~0.1\%; see Paper~IV).

%To recover radio quasars that have offsets of more than~2.0$''$,
%we separately identify all objects with a greater than
%3$\sigma$ detection of FIRST flux at the optical position (1596 sources).
%For these objects as well as those with a FIRST catalog match within
%2$''$, we perform a second FIRST catalog search with 30$''$ matching radius
%to identify possible radio lobes associated with the quasar, finding
%such matches for 2440 sources.

Matches with the {\it ROSAT} All-Sky Survey Bright and
Faint Source Catalogs (Voges et al. 1999, 2000)
were made with a maximum allowed positional offset
of~30$''$; this is similar to the 27.5$''$ positional coincidence
required for the SDSS
{\it ROSAT} target selection code (Anderson et al.~2003).
During the ROSAT all-sky survey phase, some regions in the sky received
little or no exposure due to a malfunction of the satellite's
attitude measuring and control system. These regions of the sky
were reobserved -- from 1997 February~4 to February~26 -- by a set of 783
pointings (the satellite had lost its capability to perform survey mode
operations) with typical exposures of 1500~s. Approximately 2900 X-ray
source detections were obtained in these fields; 2708 are unique objects
and are not contained in the bright and faint source catalogues of ROSAT.
These supplemental sources were included in our {\it ROSAT} matching.

The catalog contains~5377 RASS matches; approximately~1.3\% are expected
to be false identifications (Paper~IV).
The SDSS-RASS offsets for these matches are
presented in the upper right panel of Figure~6; the solid curve, which is
the Rayleigh distribution for a $1\sigma$
positional error of~10.89$''$ (fit using all the points), matches the data
quite well.

Infrared photometric information for the quasars was provided by the
Two Micron All Sky Survey (2MASS; Skrutskie et al. 2006).  The SDSS quasar
positions were compared to the 2MASS All-Sky and ``6x" point source
catalogs using a matching radius of~2$\arcsec$.  (The 6x catalog covered
a smaller area of sky six times deeper than the main survey.)
If matches were found
in both the All-Sky and 6x catalogs, the information in the latter was
selected for the quasar catalog.  In the rare
(approximately~1.5\%) cases where this approach found more than one
2MASS match for an SDSS quasar in a given catalog,
the 2MASS source with the smallest separation was adopted.  This
procedure identified 2MASS photometry for 13,930 of the quasars.
This approach is similar to, but slightly more extensive, than the 2MASS
matching in previous SDSS quasar catalogs.

Since the
2MASS catalogs represent only the most secure
($5\sigma$) sources, and our object positions are known to
considerable accuracy, it is possible to obtain additional information by
extracting photometric data from the 2MASS images themselves.
We have performed standard aperture photometry ($2''$~radius) on the 2MASS
images, following the
procedure outlined in Gallagher et al. (2008), and report all
detections larger than 2$\sigma$ in the quasar catalog.  (The ``2$\sigma$"
objects have the 2MASS flag set to ``2" in the quasar catalog.)
A total of 39,634 quasars had
$2\sigma$ detections in at least one 2MASS band, bringing the total
number of objects in the catalog with infrared photometry to~53,564.
Of the 1171 quasars with \hbox{$i < 17$}, almost all (1167)
have a 2MASS $J$-band detection.
The offsets between the SDSS and 2MASS positions are shown in the lower
left panel of Figure~6.    Only the matches with the 2MASS catalog are
displayed in the graph.

%\subsection{Spectroscopic Target Information}
%
%About two-thirds of the catalog entries
%were selected based on the SDSS
%quasar selection criteria (either a low-redshift or high-redshift candidate,
%or both) described in GTR02.  Slightly more than
%half of the quasars
%in the catalog are serendipity-flagged candidates,
%which is also primarily an ``unusual
%color" algorithm; about one-fifth of the catalog was selected by
%the serendipity criteria alone.

The mechanical constraint
that SDSS spectroscopic fibers must be separated by~55$''$ on a given plate
makes it difficult for the spectroscopic survey to confirm close pairs
of quasars.  In regions that
are covered by more than one plate, however, it is possible
to obtain spectra of both components of a close pair.
There are~449 pairs of quasars in the catalog with
angular separation less than
an arcminute, of which 46 have angular separations less than 20$''$.
Most of the pairs are chance superpositions, but there
are many sets whose components have similar redshifts,
suggesting that the quasars may be physically
associated (e.g., Hennawi et al.~2006).
The typical uncertainty in the measured value of the redshift
difference between two quasars is~$\approx$~0.015;
the catalog contains 27 quasar pairs
with separations of less than an arcminute and \hbox{with $\Delta z < 0.02$.}
These pairs, which are excellent candidates for binary quasars, are listed
in Table~4.  One of the pairs, \hbox{SDSS J140012.77+313454.1}
and \hbox{SDSS J140012.85+313452.7,} at a redshift of~3.3 and a separation
of~1.7$''$, is known to be
a gravitational lens (Inada et al.~2008).

\section{Discussion}

The DR5 database is entirely contained in that of DR7, but there are
181 quasars from Paper~IV (0.17\%) that do not have a
counterpart within~1.0$''$ of a DR7 quasar.  (This fraction is similar
to the 0.15\% dropped between Papers~III and~IV.)  Of the 181, there
are ten that are included in the DR7 list that match at larger
offsets; eight have matches within~2$''$
and the other two have separations of 2.0$''$ and~5.3$''$.  These ten
separations are much larger than one would expect to find based on the
0.1$''$ rms scatter in the positions; the large coordinate offsets
presumably arise from different versions of the imaging pipeline
software (in particular the deblender).

The majority of the
dropped objects are low-redshift, low-luminosity sources
($\approx$~80\% have redshifts below one).
Most of the changes were introduced because 1)~the spectra did not appear
to satisfy our quasar criteria (primarily the minimum line width limit)
based upon the PCA analysis or 2)~small
changes in the photometric measurements dropped the luminosity below our
absolute magnitude criterion.

Although the quasars in the catalog have been selected using a variety of
techniques (optical colors, radio properties, X-ray emission), the information
contained in the catalog allows one to easily construct well-defined
subsamples.  For example, to construct an optically-selected set of quasars
(e.g., Strateva et al. 2005),
one can either require that the low-redshift or high-redshift
target selection flag is set, or, alternatively, require that the object
was not chosen solely by radio/X-ray criteria (depending on whether or not
you wish to have AGN found by the serendipity, galaxy, or star algorithms),
using the quantities in \hbox{columns 55--61} in the catalog.

The primary purpose of this catalog is to compile all of the quasars
discovered/recovered by the SDSS survey in a complete and robust
manner, allowing the general user to avoid the pitfalls that can arise
from data mining such a large repository.  All of the spectra have been
inspected by eye, frequently by more than one of the authors.  It may
be that other users desire a more inclusive list of AGNs; for example,
by relaxing our cuts on luminosity or line width.  In creating an
expanded AGN list, it is important to consider two important issues in
terms of the compilation and statistical analysis of such samples.

First, it is tempting to eliminate the labor-intensive visual
examination stage and rely on the {\tt zconf} flag as a means of
restricting the AGN sample to the most robust objects.  However, {\tt zconf}
is not a good measure of the reliability of quasar redshifts:
it depends strongly on redshift, as different emission lines enter and
leave the SDSS spectral coverage.  For example, {\tt zconf} drops
dramatically in the mean from $z \sim 0.7$ to $z \sim 0.9$ as the
H$\beta$ feature leaves the SDSS spectral bandpass. The left panel of Figure~7
shows {\tt zconf} as a function of redshift for bona-fide quasars
whose spectra have been confirmed by eye.  The red histogram in the
right panel in Figure~7 demonstrates the result of applying an
arbitrary \hbox{{\tt zconf} $>0.95$} cut, independent of redshift, to
the DR7 quasar sample.  The redshift dependence of {\tt zconf}
introduces an artificial apparent periodicity in the redshift
distribution.

The second issue has to do with the effects of emission lines on
quasar photometry. The SDSS quasar selection will include
intrinsically fainter objects whenever a strong emission line enters
the $i$ bandpass, making the quasar appear brighter than the same
quasar at a redshift where the observed $i$ filter covers only
continuum emission (Richards et al. 2006).
When we
restrict the sample to $i < 19.1$ and correct for the emission line
k-correction (green histogram in Figure~7), the redshift distribution
of the DR7 quasars is quite smooth.

\section{Summary}

This is the final edition of the Sloan Digital Sky Survey~I and~II
Quasar Catalog,
which was compiled from observations taken
between~1999 and~2008.  One of the original stated goals of the SDSS was
to obtain spectra of 100,000 quasars, a total that has been met with this
catalog.
The lower right panel in Figure~6 charts the progress of the SDSS Quasar
Survey, denoted by the number of spectroscopically-confirmed quasars
as a function of time, over the duration of the project.

In the summer of 2008 the third phase of the SDSS, denoted as SDSS-III,
began; this survey is planned to be in operation for six years.  One of
the components of SDSS-III, the Baryonic Oscillation Spectroscopic Survey
(BOSS; Schlegel, White, \& Eisenstein 2009), will use improved spectrographs
on the SDSS telescope to obtain spectra of 1.5 million luminous galaxies
and over 100,000 \hbox{$z \approx 2.5$} quasars.
We plan to continue
our policy of timely releases of high-quality quasar catalogs throughout the
SDSS-III era.

\acknowledgments

We would like to thank Paul Hewett for his very helpful comments on an
early version of the DR7 quasar catalog.
This work was supported in part by National Science Foundation grants
AST-0607634 (DPS, DVB, NPR), AST-0307384~(XF), and
AST-0707266 (MAS and YS), and by
NASA LTSA grant NAG5-13035 (WNB, DPS).
GTR was supported in part by a \hbox{Alfred P. Sloan}
Foundation Fellowship, and
PBH acknowledges support by NSERC.
XF acknowledges support from an \hbox{Alfred P. Sloan} Fellowship and
a David and Lucile Packard Fellowship in Science and Engineering.
%SJ was supported by the Max-Planck-Gesellschaft (MPI f\"ur Astronomie) through
%an Otto Hahn fellowship.
%CS was supported by the U.S. Department of Energy under contract
%\hbox{DE-AC02-76CH03000.}

Funding for the SDSS and SDSS-II has been
provided by the Alfred P.~Sloan Foundation,
the Participating Institutions,
the National Science Foundation,
the U.S. Department of Energy,
the National Aeronautics and Space Administration,
the Japanese Monbukagakusho,
the Max Planck Society,
and the Higher Education Funding Council for England.
The SDSS Web site \hbox{is {\tt http://www.sdss.org/}.}

The SDSS is managed by the Astrophysical Research Consortium
(ARC) for the Participating Institutions.  The participating
institutions are
the American Museum of Natural History,
Astrophysical Institute of Potsdam,
University of Basel,
Cambridge University,
Case Western Reserve University,
University of Chicago,
Drexel University,
Fermilab,
the Institute for Advanced Study,
the Japan Participation Group,
Johns Hopkins University,
the Joint Institute for Nuclear Astrophysics,
the Kavli Institute for Particle Astrophysics and Cosmology,
the Korean Scientist Group,
the Chinese Academy of Sciences (LAMOST),
Los Alamos National Laboratory,
the Max-Planck-Institute for Astronomy (MPIA),
the Max-Planck-Institute for Astrophysics (MPA),
New Mexico State University,
Ohio State University,
University of Pittsburgh,
University of Portsmouth,
Princeton University,
the United States Naval Observatory,
and the University of Washington.

This research has made use of 1)~the NASA/IPAC Extragalactic Database (NED)
which is operated by the Jet Propulsion Laboratory, California Institute
of Technology, under contract with the National Aeronautics and Space
Administration, and 2)~data products from the Two Micron All Sky
Survey, which is a joint project of the University of
Massachusetts and the Infrared Processing and Analysis Center/California
Institute of Technology, funded by the National Aeronautics
and Space Administration and the National Science Foundation.

%The Hobby-Eberly Telescope (HET) is a joint project of the University of Texas
%at Austin,
%the Pennsylvania State University,  Stanford University,
%Ludwig-Maximillians-Universit\"at M\"unchen, and Georg-August-Universit\"at
%G\"ottingen.  The HET is named in honor of its principal benefactors,
%William P. Hobby and Robert E. Eberly.  The Marcario Low-Resolution
%Spectrograph is named for Mike Marcario of High Lonesome Optics, who
%fabricated several optics for the instrument but died before its completion;
%it is a joint project of the Hobby-Eberly Telescope partnership and the
%Instituto de Astronom\'{\i}a de la Universidad Nacional
%Aut\'onoma de M\'exico.

\appendix
\section{Quasars Lacking Catalog Archive Server Photometry}

We required that every quasar in our catalog had photometry available
in the CAS (the photometry was required to determine if the object met
our luminosity criterion).
There were 6383 objects with spectra in the quasar master list without
photometry, of which 207 are genuine quasars.
A few of these quasars were missed
due to archiving errors in matching the BEST, TARGET and
spectroscopic data; the remainder are drawn from four ``special''
plates (see the DR4 paper, Adelman-McCarthy et al.~2006) whose photometry was
not entered into the CAS in a consistent manner.  Three of those plates
(1468, 1471, and 1472) targetted objects with quasar colors from
fields in the vicinity of M31 (Adelman-McCarthy et al.~2007); these
quasars can be used to study absorption from gas in M31.  A fourth
plate, 797, selected F star candidates from a field in Stripe 82, but
the matching to the photometry was never done properly.

The data for these
objects are included in Table 5, which includes a subset of the
data columns of the primary catalog.  With two exceptions,
these objects all have listed PSF photometry drawn from the Data Archive
Server.
The redshifts range from 0.148 to 3.18, with a median of 1.5, i.e.,
very similar to that of the primary catalog quasars.
These quasars are not included in the summary figures in the paper.
Table 5 contains the following information for each quasar:
SDSS name, J2000 right ascension and declination (degrees), redshift,
$ugriz$ PSF magnitudes and errors (uncorrected for Galactic extinction),
Galactic extinction in the $u$ filter
(magnitudes), morphological type (0 = star, 1 = galaxy), Primary Target
flag, photometric run number, rerun number, camera column, field and
object~ID, and the MJD, plate, and fiber values for the spectrum.

\newpage

\begin{figure}
\includegraphics[angle=270,scale=0.70,viewport= 0 0 500 700]{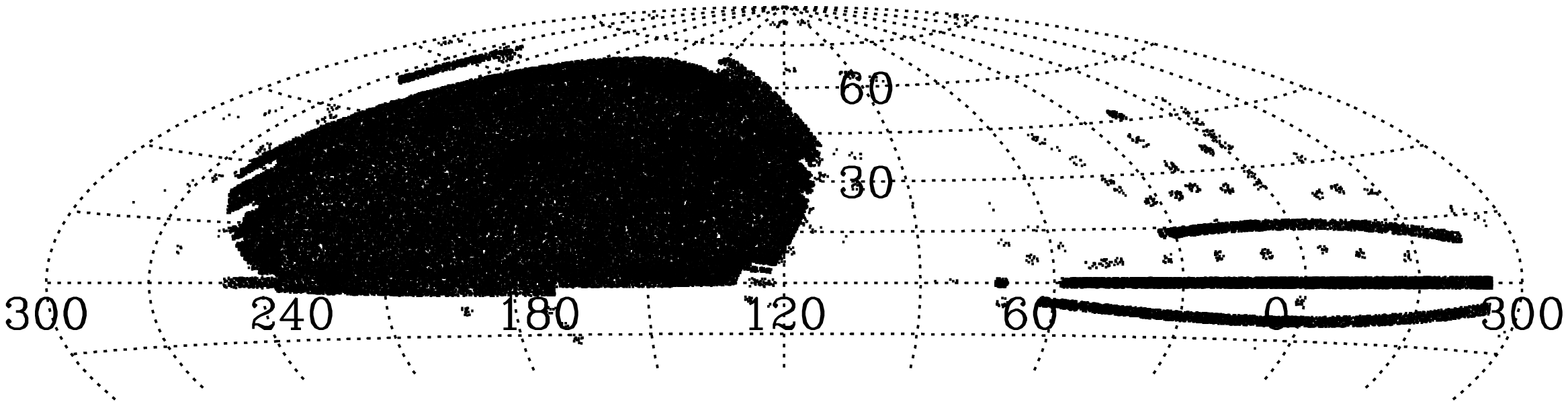}
%\plotone{f1.ps}{5.0in}{90.0}{70.0}{70.0}{270.0}{0.0}
\caption{
The sky distribution of the 105,783 quasars in the catalog in J2000
equatorial coordinates.  The primary
components of the dataset are those associated with the Legacy
Survey (North Galactic Cap, the large contiguous area, and the South
Galactic Cap, three narrow stripes aligned with the Celestial
Equator) and SEGUE (the points lying off the stripes in the South
Galactic Cap).
\label{Figure 1 }
}
\end{figure}

\clearpage

\begin{figure}
\includegraphics[angle=0,scale=0.70,viewport= 0 0 500 700]{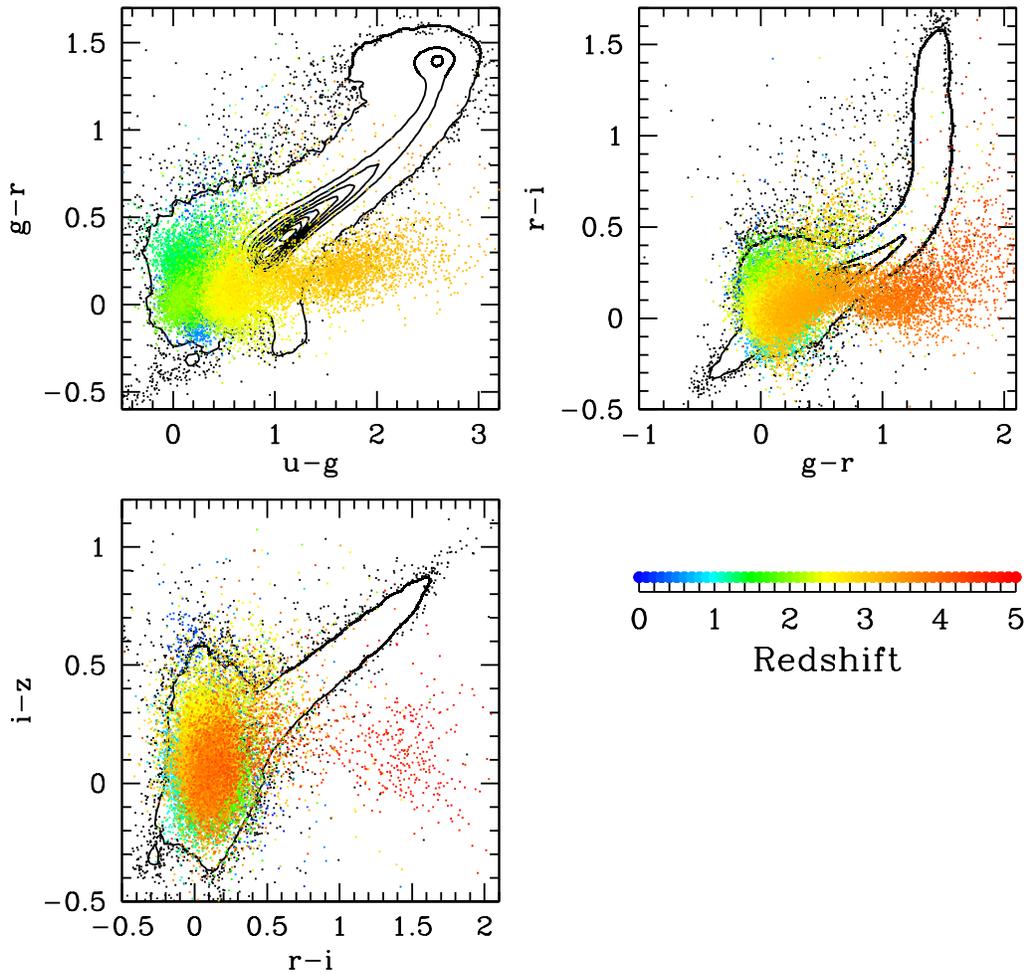}
%\plotone{f1.ps}{5.0in}{90.0}{70.0}{70.0}{270.0}{0.0}
\caption{
The distribution of unresolved sources (black dots and contours) and
quasars (color dots) in SDSS $ugriz$
color-color space.  The quasars are color-coded by redshift.  For clarity
we only show 10\% of the $z<2.2$ quasars in the figure.
\label{Figure 2 }
}
\end{figure}

\clearpage

\begin{figure}
\includegraphics[angle=90,scale=0.70,viewport= 0 0 500 700]{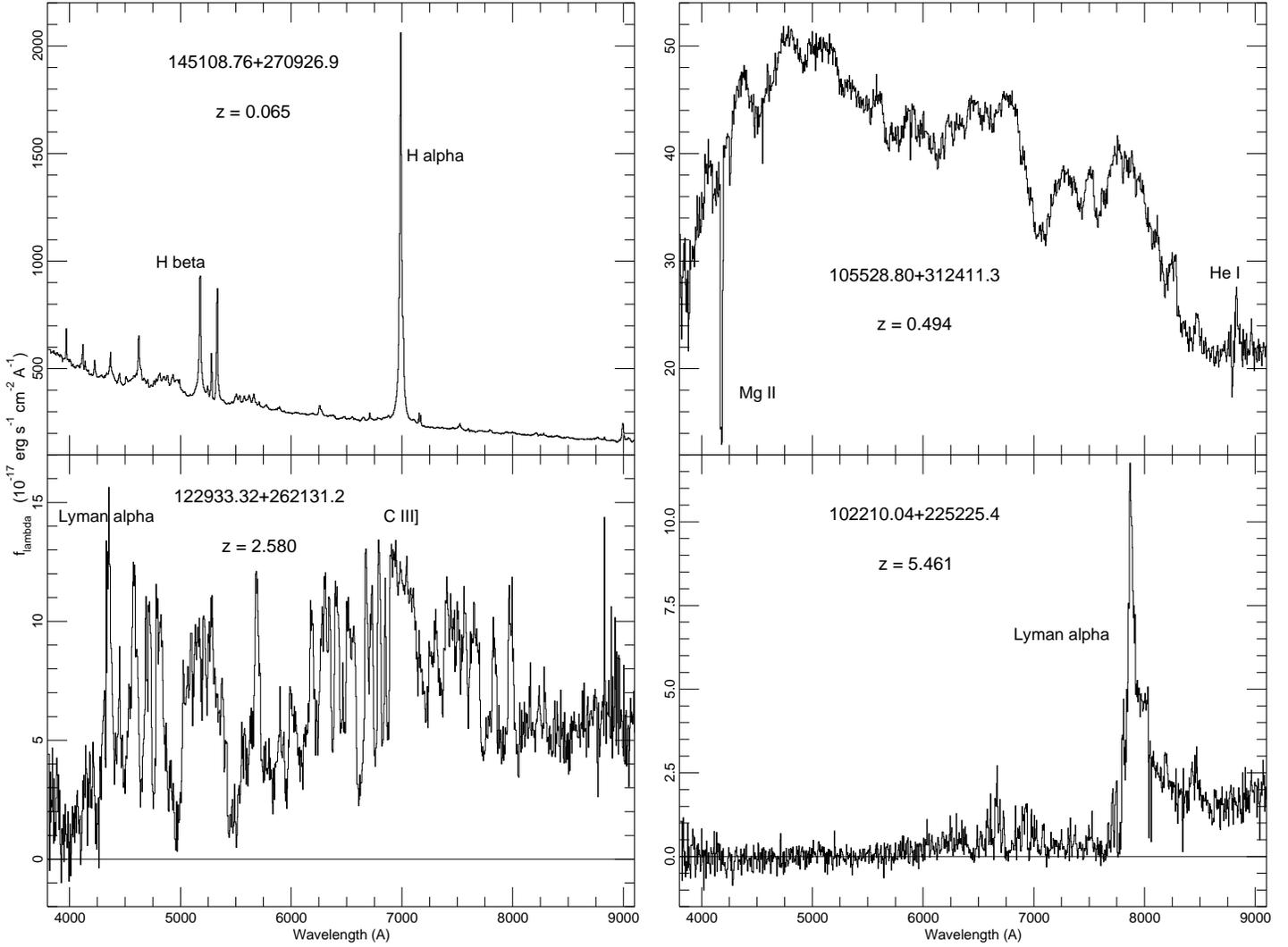}
%\plotone{f1.ps}{5.0in}{90.0}{70.0}{70.0}{270.0}{0.0}
\caption{
SDSS spectra of four quasars that have been obtained
since the DR5 catalog.
The spectral
resolution of the data ranges from 1800 to~2100.
The data have been rebinned \hbox{to 5 \AA\ pixel$^{-1}$}
for display purposes.  The panels display (clockwise, starting in upper
left) 1)~the lowest redshift quasar in the catalog (previously known;
\hbox{PG 1448+273,} Schmidt \& Green~1983), 2)~a quasar with unusually strong
Fe emission (previously known; \hbox{2MASS J10552880+3124112,}
Barkhouse \& Hall~2001),
3)~the highest redshift quasar in the catalog (note the nearly complete
absorption below the Lyman~$\alpha$ emission line and the narrow C~IV
absorption
just longward of the Lyman~$\alpha$ emission line), and 4)~a striking
example of an extreme BAL quasar.
\label{Figure 3 }
}
\end{figure}

\begin{figure}
\includegraphics[angle=0,scale=0.90]{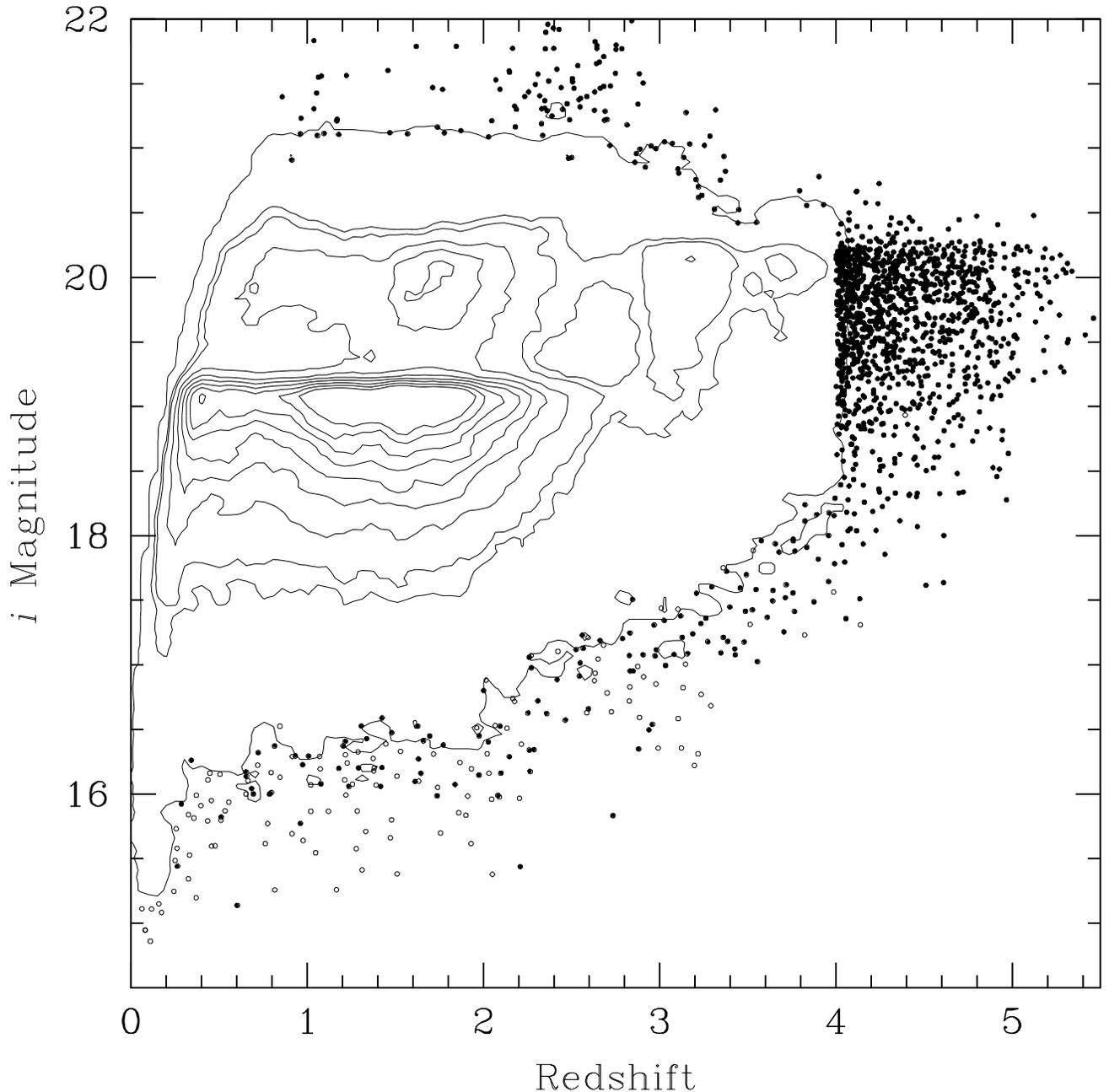}
%\plotone{f2.ps}{5.0in}{0.0}{90.0}{70.0}{270.0}{0.0}
%\plotfiddle{dr3qsofiz.ps}{6.5in}{0.0}{90.0}{90.0}{-270.0}{-100.0}
\figcaption{
The observed~$i$ magnitude as a function of redshift for the~105,783
objects in the catalog.  Open circles indicate quasars that
were recovered, but not discovered, by
the SDSS.
The distribution is represented by a set of linear contours when the
density of points in this two-dimensional space would cause
the points to overlap.
The steep gradients at \hbox{$i \approx 19.1$} and
\hbox{$i \approx 20.2$} are due to the flux limit for the
targeted low and high redshift parts of the survey;
the dip in the counts at $z$~$\approx$~2.7 arises
because of the high incompleteness of the SDSS Quasar Survey at redshifts
between 2.5 and~3.0 (also see Figure~5).
\label{Figure 4 }
}
\end{figure}

\begin{figure}
\includegraphics[angle=90,scale=0.70,viewport= 0 0 500 700]{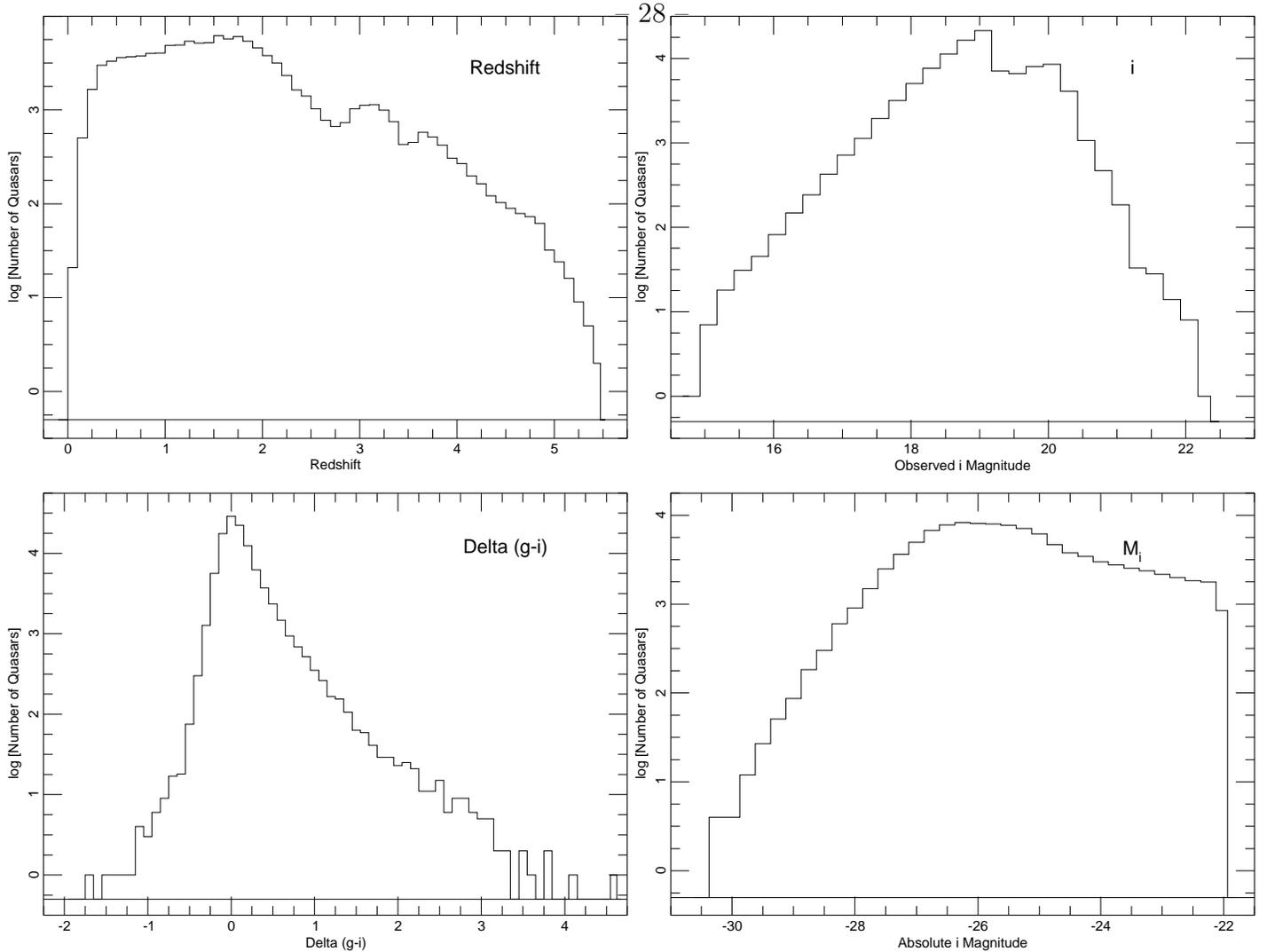}
%\plotfiddle{dr3qsofmatch.ps}{5.0in}{90.0}{70.0}{70.0}{280.0}{0.0}
\figcaption{
The four panels display the distributions of the catalog quasars in redshift,
observed~$i$~magnitude, $\Delta (g-i)$, and absolute magnitude. The two left
panels have bin sizes of~0.1; the two panels on the right have magnitude
intervals of~0.25.  Logarithmic scales have been adopted because of the
enormous dynamic range in all of the histograms.
If there are zero points in a bin, the logarithm
of the counts is set to $-0.3$ (horizonal lines in each panel).
The redshifts range from~0.065 to~5.46.
The dips at redshifts of~2.7 and~3.5
are caused by the reduced completeness of the selection algorithm at these
redshifts.
The $i$ magnitude in the upper right panel is not corrected for Galactic
absorption. The sharp
drop that occurs at magnitudes slightly fainter than~19 is due to the
flux limit for the low-redshift targeted part of the survey.
The SDSS Quasar survey has
a bright limit of \hbox{$i \approx 15.0$} imposed by the need to avoid
saturation in the spectroscopic observations.  The distribution of
the $\Delta(g-i)$ colors shows strong asymmetry, with a tail of objects
extending to the red.
\label{Figure 5 }
}
\end{figure}
%
%\begin{figure}
%%\includegraphics[angle=0,scale=0.90]{dr7qfczr.eps}
%\includegraphics[angle=0,scale=0.90]{fig6.ps}
%%\plotfiddle{dr3qsofz5.ps}{5.0in}{90.0}{70.0}{70.0}{275.0}{0.0}
%\figcaption{
%The quasar color-redshift relation for the DR7 quasars (photometry corrected
%for Galactic extinction).  Contours are used to represent the distribution
%when the density of points would cause the points to overlap.  The panels
%present the four standard SDSS colors; the dashed gray lines are the modal
%relations presented in Table~5.  The influence of emission lines on the colors
%is readily apparent.
%The tightness of the correlations breaks down when the Lyman~$\alpha$ forest
%region dominates the bluer of the two passbands (e.g., above redshifts
%of~2.2 in the $(u-g)$ relation).
%\label{Figure 6 }
%}
%\end{figure}

\begin{figure}
\includegraphics[angle=90,scale=0.70,viewport= 0 0 500 700]{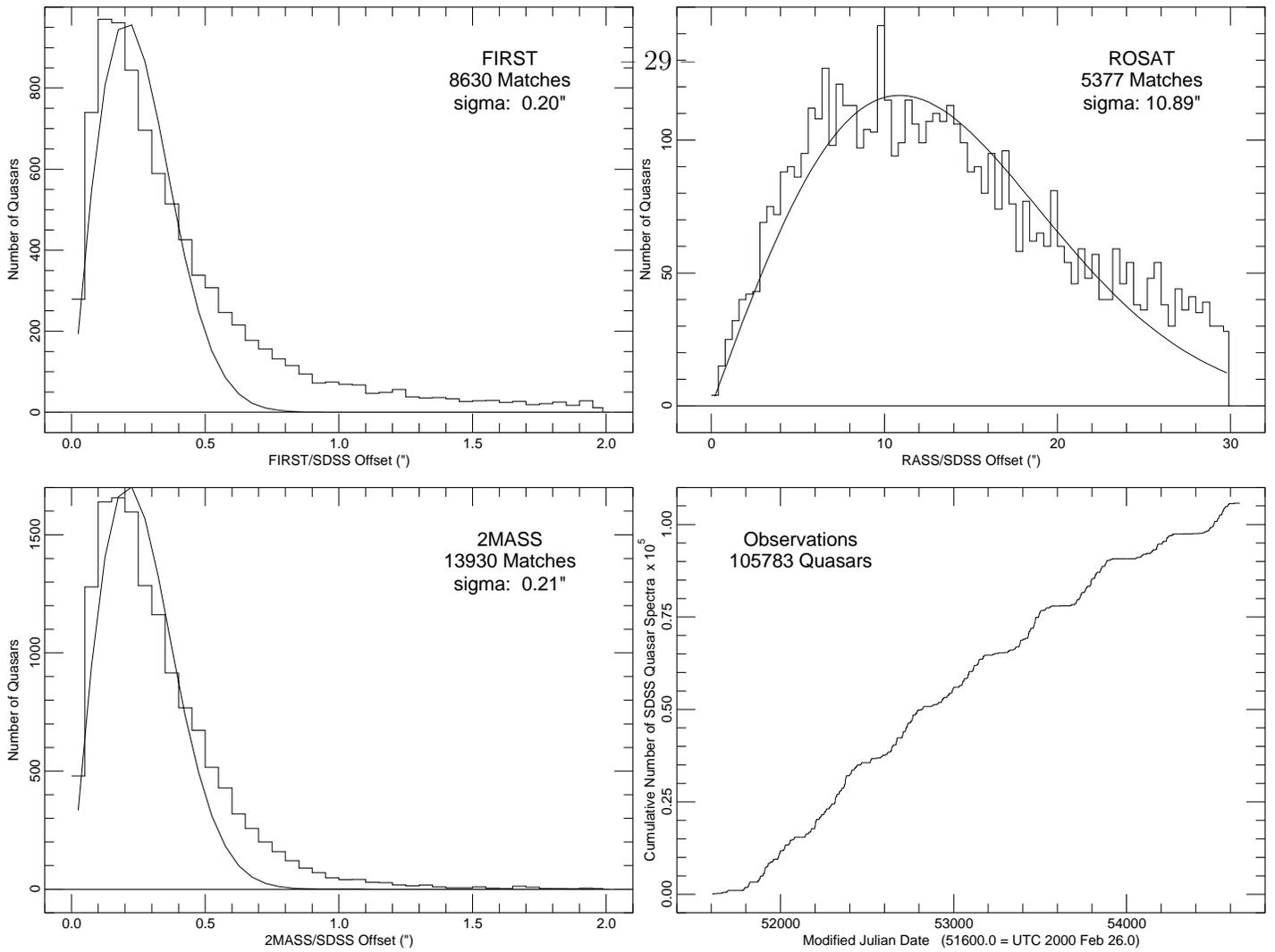}
%\plotfiddle{dr3qsofmatch.ps}{5.0in}{90.0}{70.0}{70.0}{280.0}{0.0}
\figcaption{
The upper row and lower left panels show the distribution of the angular
offsets between the SDSS positions and those of the FIRST, ROSAT, and 2MASS
coordinates, respectively.
The matching radii for the surveys
were~$2.0''$ (FIRST), $30.0''$ (ROSAT), and~$2.0''$ (2MASS).
The smooth curves in each of these panels
is the expected
distribution for a set of matches if the offsets between the objects are
described by a Rayleigh distribution; the dispersions of the fit are
given in each panel.  Only points with separations of less than~$1.0''$
were used in fitting the Rayleigh distribution in the FIRST and
2MASS matching.
In the 2MASS case, only points
that were actual matches to the published 2MASS catalog are displayed.
The Rayleigh distribution is a poor fit to the FIRST and 2MASS matches,
as both data sets have a much higher than expected number of matches
at radii larger than~$\approx$~0.5$''$.
The lower right panel displays the cumulative number of DR7 quasars as a
function of time; the observations occurred between
February~2000 and July~2008.  The roughly periodic structure
in the curve is produced by the yearly summer maintence schedule and,
late in the survey, the limited number of quasar observations during the
Fall once that portion of the Legacy Survey spectroscopy was completed.
\label{Figure 6 }
}
\end{figure}

\clearpage

\begin{figure}
%\plottwo{dr7qfzconf.eps}{dr7qfzchist.eps}
\plottwo{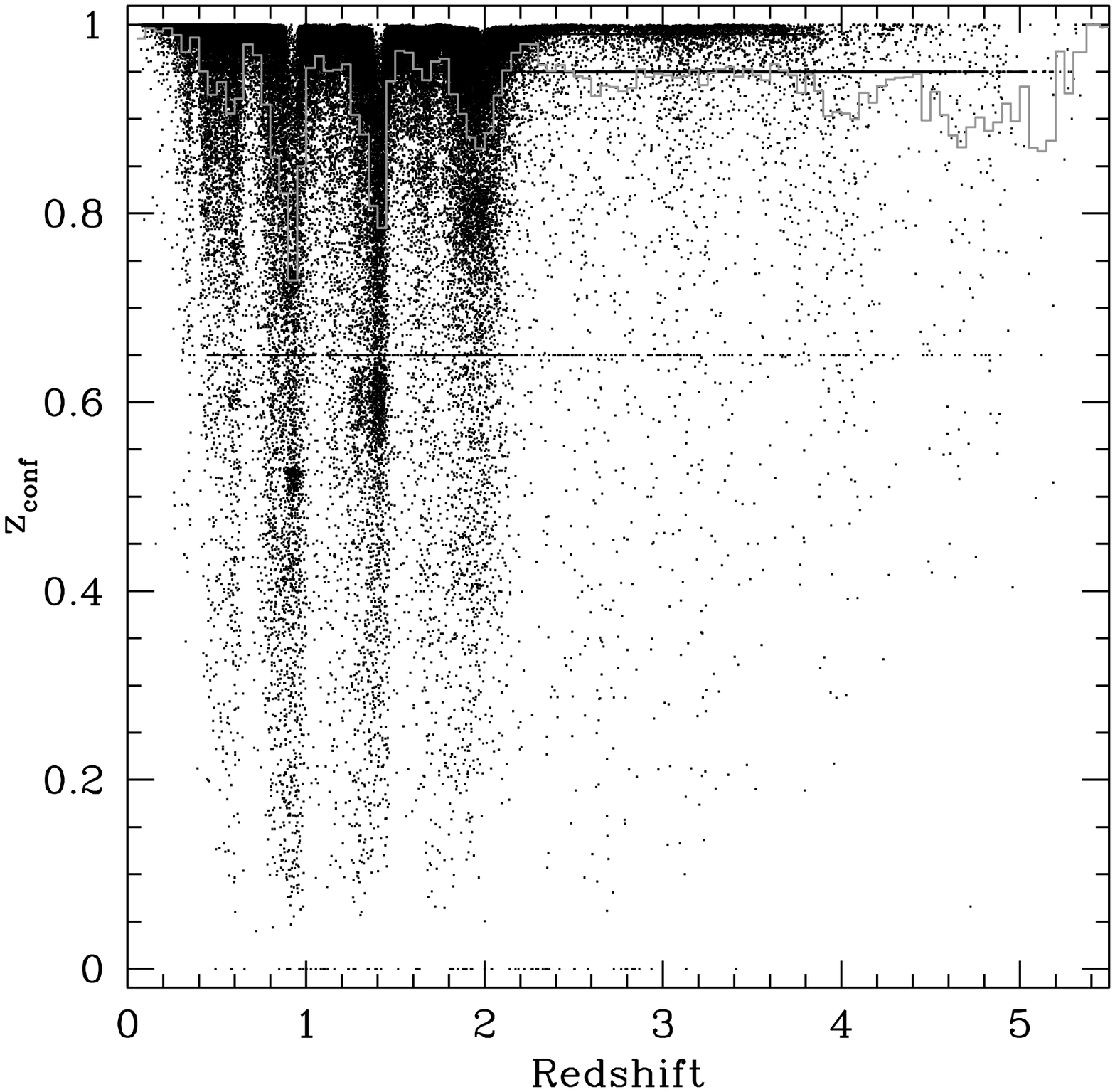}{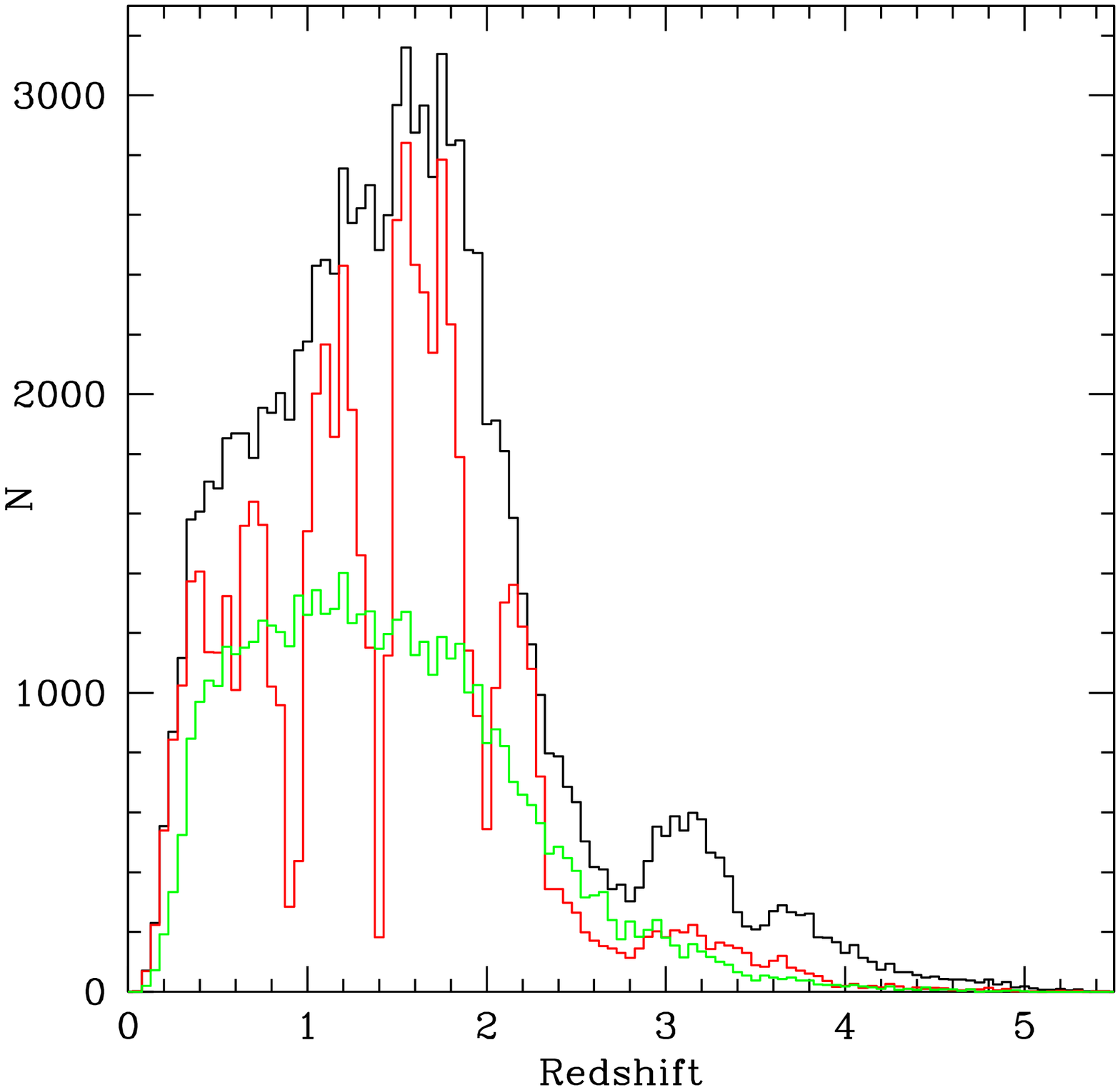}
\figcaption{
Left panel:
The distribution in the {\tt zconf}-redshift plane for the
DR7 quasars.  Black points indicate individual objects; the
gray line shows the moving average in redshift bins of 0.05.  Note the
banding structure produced by the movement of strong emission lines
through the SDSS spectrograph bandpass.  A large number of low {\tt zconf}
values occur at the redshifts where there are fewer strong
emission features in the SDSS spectra (see text).
Right panel: The
black histogram shows the raw redshift distribution of the DR7 quasars
in the catalog.  The red histogram displays the redshift distribution
that results from applying a cut of \hbox{{\tt zconf} = 0.95} on the points
in the left-hand
panel; this selection
produces an unphysical redshift distribution.  This is not an issue
for quasars in this catalog, but must be considered in any attempt to
extend the catalog (e.g., to fainter quasars) for the purpose of
statistical analyses.  The green histogram corrects for the
emission-line Malmquist bias and reveals the intrinsic redshift
distribution for a fixed continuum luminosity (see text).
\label{Figure 7 }
}
\end{figure}

\clearpage

%-------- Tables Start Here ----------------------

\begin{deluxetable}{rcl}
%\tabletypesize{\small}
%\rotate
\tablewidth{0pt}
\tablecaption {SDSS DR7 Quasar Catalog Format}
\tablehead{
\colhead{Column} &
\colhead{Format} &
\colhead{Description}
}

\startdata
   1  &  A18  &
SDSS DR7 Designation   \ \ \ \ hhmmss.ss+ddmmss.s  \ \ \ (J2000) \\
   2  &  F11.6  &   Right Ascension in decimal degrees (J2000) \\
   3  &  F11.6  &   Declination in decimal degrees (J2000) \\
   4  &  F7.4 &   Redshift \\
%   5  &   I2  &   Database query that recovered object (1, 2, 3, or 4) \\
   5  &   F7.3  &    BEST PSF $u$ magnitude
(not corrected for Galactic extinction) \\
   6  &   F6.3  &    Error in BEST PSF $u$ magnitude \\
   7  &   F7.3  &    BEST PSF $g$ magnitude
(not corrected for Galactic extinction) \\
   8  &   F6.3  &    Error in BEST PSF $g$ magnitude \\
   9  &   F7.3  &    BEST PSF $r$ magnitude
(not corrected for Galactic extinction) \\
  10  &   F6.3  &    Error in BEST PSF $r$ magnitude \\
  11  &   F7.3  &    BEST PSF $i$ magnitude
(not corrected for Galactic extinction) \\
  12  &   F6.3  &    Error in BEST PSF $i$ magnitude \\
  13  &   F7.3  &    BEST PSF $z$ magnitude
(not corrected for Galactic extinction) \\
  14  &   F6.3  &    Error in BEST PSF $z$ magnitude \\
  15  &   F7.3  &    $u$ band Galactic extinction (from Schlegel et al 1998) \\
%  17  &   F7.3  &   Rest wavelength (\AA ) of emission line with largest
%velocity FWHM \\
%  18  &   F7.3  &   FWHM (km s$^{-1}$) of emission line with largest
%velocity FWHM \\
  16  &   F7.3  &    $\log N_H$  (logarithm of Galactic H I column
density in cm$^{-2}$) \\
  17  &   F7.3  &    FIRST peak flux density at 20 cm expressed as AB
magnitude; \\
& & \ \ \ \ \ 0.0 is no detection, $-1.0$ source is not in FIRST area \\
  18  &   F8.3  &    S/N of FIRST flux density \\
  19  &   F7.3  &    SDSS-FIRST separation in arc seconds \\
%  20  &   I3    &   $> 3\sigma$ FIRST flux at optical position
%but no FIRST counterpart within 2$''$ (0 or 1) \\
%  29  &   I3    &   FIRST source located 2$''$-30$''$ from optical position
%(0 or 1) \\
  20  &   F8.3  &   log RASS full band count rate (counts s$^{-1}$);
$-9.0$ is no detection \\
  21  &   F7.3  &   S/N of RASS count rate \\
  22  &   F7.3  &   SDSS-RASS separation in arc seconds \\
  23  &   F7.3  &   $J$ magnitude (2MASS);
0.0 indicates no 2MASS detection \\
  24  &   F6.3  &   Error in $J$ magnitude \\
  25  &   F7.3  &   $H$ magnitude (2MASS);
0.0 indicates no 2MASS detection \\
  26  &   F6.3  &   Error in $H$ magnitude \\
  27  &   F7.3  &   $K$ magnitude (2MASS);
0.0 indicates no 2MASS detection \\
  28  &   F6.3  &   Error in $K$ magnitude \\
  29  &   F7.3  &   SDSS-2MASS separation in arc seconds \\
  30  &   I3    &   2MASS Flag = 9$*J$ flag + 3$*H$ flag + $K$ flag \\
& & \ \ \ \ Filter flags \ \ 0 = no detection,
1 = catalog match, 2 = new photometry \\
  31  &   F8.3  &   $M_{i}$ ($H_0$ = 70 km s$^{-1}$ Mpc$^{-1}$,
$\Omega_M = 0.3$, $\Omega_{\Lambda} = 0.7$, $\alpha_{\nu} = -0.5$) \\
  32  &   F7.3  &   $\Delta(g-i) = (g-i) - \langle (g-i)
\rangle_{\rm redshift}$ (Galactic extinction corrected)  \\
  33  &   I3  &   SDSS Morphology flag \ \ \ 0 = point source
\ \ \ 1 = extended \\
  34  &   I3  &   SDSS Spectroscopy SCIENCEPRIMARY flag  (0 or 1) \\
  35  &   I3  &   SDSS MODE flag (blends, overlapping scans; 1, 2, or 3) \\
  36  &   I3  &   Selected with final quasar algorithm (0 or 1) \\
  37  &  I12  &   Target Selection Flag (BEST) \\
  38  &   I3  &   Low-$z$ Quasar selection flag (0 or 1) \\
  39  &   I3  &   High-$z$ Quasar selection flag (0 or 1) \\
  40  &   I3  &   FIRST selection flag (0 or 1) \\
  41  &   I3  &   {\it ROSAT} selection flag (0 or 1) \\
  42  &   I3  &   Serendipity selection flag (0 or 1) \\
  43  &   I3  &   Star selection flag (0 or 1) \\
  44  &   I3  &   Galaxy selection flag (0 or 1) \\
  45  &   I6  &   SDSS Imaging Run Number of BEST photometric measurements \\
  46  &   I6  &   Modified Julian Date of BEST imaging observation \\
  47  &   I6  &   Modified Julian Date of spectroscopic observation \\
  48  &   I5  &   Spectroscopic Plate Number \\
  49  &   I5  &   Spectroscopic Fiber Number \\
  50  &   I4  &   SDSS Photometric Processing Rerun Number \\
  51  &   I3  &   SDSS Camera Column Number (1--6) \\
  52  &   I5  &   SDSS Field Number \\
  53  &   I5  &   SDSS Object Identification Number \\
  54  &  I12  &   Target Selection Flag (TARGET) \\
  55  &   I3  &   Low-$z$ Quasar selection flag (0 or 1) \\
  56  &   I3  &   High-$z$ Quasar selection flag (0 or 1) \\
  57  &   I3  &   FIRST selection flag (0 or 1) \\
  58  &   I3  &   {\it ROSAT} selection flag (0 or 1) \\
  59  &   I3  &   Serendipity selection flag (0 or 1) \\
  60  &   I3  &   Star selection flag (0 or 1) \\
  61  &   I3  &   Galaxy selection flag (0 or 1) \\
  62  &   F7.3  &    TARGET PSF $u$ magnitude
(not corrected for Galactic extinction) \\
  63  &   F6.3  &    Error in TARGET PSF $u$ magnitude \\
  64  &   F7.3  &    TARGET PSF $g$ magnitude
(not corrected for Galactic extinction) \\
  65  &   F6.3  &    Error in TARGET PSF $g$ magnitude \\
  66  &   F7.3  &    TARGET PSF $r$ magnitude
(not corrected for Galactic extinction) \\
  67  &   F6.3  &    Error in TARGET PSF $r$ magnitude \\
  68  &   F7.3  &    TARGET PSF $i$ magnitude
(not corrected for Galactic extinction) \\
  69  &   F6.3  &    Error in TARGET PSF $i$ magnitude \\
  70  &   F7.3  &    TARGET PSF $z$ magnitude
(not corrected for Galactic extinction) \\
  71  &   F6.3  &    Error in TARGET PSF $z$ magnitude \\
  72  &   I21   & BestObjID (64 bit integer) \\
  73  &   I21   & SpecObjID (64 bit integer) \\
  74  &   1X, A25 & NED Object Name for previously known quasars \\
 & & \ \ \ ``SDSS" designates previously published SDSS object \\
\enddata

\end{deluxetable}

\clearpage

\begin{deluxetable}{crrrcrrrrrrrrrr}
\tabletypesize{\small}
\rotate
\tablewidth{0pt}
\tablecaption {The SDSS Quasar Catalog V$^{a}$}
\tablehead{
\colhead{Object (SDSS J)} &
\colhead{R.A. (deg)} &
\colhead{Dec (deg)} &
\colhead{Redshift} &
\multicolumn{2}{c}{$u$} &
\multicolumn{2}{c}{$g$} &
\multicolumn{2}{c}{$r$} &
\multicolumn{2}{c}{$i$} &
\multicolumn{2}{c}{$z$} &
\colhead{$A_u$}
}

\startdata
000006.53+003055.2 & 0.027228 &  0.515341 & 1.8246 &20.384 &0.065 &20.461 &
0.034 & 20.324 &0.038 &20.093 &0.041 &20.042 &0.121  &0.130 \\
000008.13+001634.6 & 0.033900 &  0.276301 & 1.8373 &20.242 &0.054 &20.206 &
0.024 & 19.941 &0.032 &19.485 &0.032 &19.178 &0.068  &0.161 \\
000009.26+151754.5 & 0.038604 & 15.298477 & 1.1985 &19.916 &0.042 &19.807 &
0.036 & 19.374 &0.017 &19.148 &0.023 &19.312 &0.069  &0.223 \\
000009.38+135618.4 & 0.039089 & 13.938450 & 2.2342 &19.233 &0.026 &18.886 &
0.022 & 18.427 &0.018 &18.301 &0.024 &18.084 &0.033  &0.322 \\
000009.42$-$102751.9 & 0.039271 &$-$10.464426 &1.8449 &19.242 &0.036 &19.019 &
0.027 & 18.966 &0.021 &18.775 &0.018 &18.705 &0.047  &0.188 \\
\enddata

\tablenotetext{a}{
Table 2 is presented in its entirety in the electronic edition of the
Astronomical Journal.  A portion is shown here for guidance regarding
its form and content.  The full catalog
contains~74 columns of information
on~105,783 quasars.}

\end{deluxetable}

\clearpage

\begin{deluxetable}{lrrrr}
%\tabletypesize{\small}
\tablewidth{0pt}
\tablecaption {Spectroscopic Target Selection}
\tablehead{
\colhead{} &
\colhead{TARGET} &
\colhead{TARGET} &
\colhead{BEST} &
\colhead{BEST} \\
\colhead{} &
\colhead{} &
\colhead{Sole} &
\colhead{} &
\colhead{Sole} \\
\colhead{Class} &
\colhead{Selected} &
\colhead{Selection} &
\colhead{Selected} &
\colhead{Selection}
}

\startdata
Low-$z$      & 68478 &   22861 &      65823 &   20793 \\
High-$z$     & 24088 &    7789 &      24322 &    6745 \\
FIRST        &  5142 &     336 &       5261 &     319 \\
ROSAT        &  6302 &     467 &       6403 &     577 \\
Serendipity  & 56095 &   19660 &      54478 &   19333 \\
Star         &  2338 &     268 &       1194 &     242 \\
Galaxy       &   687 &     122 &        754 &     105 \\
None of Above &  5992 & & 10357 \\
\enddata

\end{deluxetable}

\clearpage

\begin{deluxetable}{ccccccr}
%\tabletypesize{\small}
\tablewidth{0pt}
\tablecaption {Candidate Binary Quasars$^{a}$}
\tablehead{
\colhead{Quasar 1} &
\colhead{Quasar 2} &
\colhead{$i_1$} &
\colhead{$i_2$} &
\colhead{$z_1$} &
\colhead{$z_2$} &
\colhead{$\Delta \theta$ $''$}
}

\startdata
001201.87+005259.7 & 001202.35+005313.9 & 20.0 & 20.2 &
 1.652 & 1.637 &  16.0 \cr
011757.99+002104.1 & 011758.83+002021.4 & 19.9 & 17.9 &
 0.612 & 0.613 &  44.5 \cr
014110.40+003107.1 & 014111.62+003145.8 & 20.4 & 20.0 &
 1.875 & 1.884 &  42.9 \cr
024511.93$-$011317.5 & 024512.12$-$011314.0 & 20.2 & 19.4 &
 2.462 & 2.460 &   4.5 \cr
025813.65$-$000326.4 & 025815.54$-$000334.2 & 20.0 & 19.0 &
 1.318 & 1.320 &  29.4 \cr
025959.68+004813.6 & 030000.57+004828.0 & 19.5 & 16.5 &
 0.893 & 0.900 &  19.6 \cr
035053.04$-$003200.1 & 035053.29$-$003114.7 & 19.5 & 19.3 &
 1.993 & 1.995 &  45.5 \cr
074336.85+205512.0 & 074337.28+205437.1 & 19.9 & 19.7 &
 1.570 & 1.565 &  35.5 \cr
074759.02+431805.3 & 074759.65+431811.4 & 19.0 & 19.2 &
 0.501 & 0.501 &   9.2 \cr
075715.09+321240.7 & 075718.02+321320.7 & 20.6 & 18.4 &
 1.446 & 1.458 &  54.6 \cr
082439.83+235720.3 & 082440.61+235709.9 & 18.9 & 18.9 &
 0.536 & 0.536 &  14.9 \cr
085625.63+511137.0 & 085626.71+511117.8 & 18.5 & 19.2 &
 0.543 & 0.542 &  21.8 \cr
090923.12+000204.0 & 090924.01+000211.0 & 20.0 & 16.4 &
 1.882 & 1.866 &  15.0 \cr
094208.81+310736.1 & 094212.95+310745.5 & 19.6 & 20.2 &
 1.708 & 1.724 &  53.9 \cr
095556.37+061642.4 & 095559.02+061701.8 & 17.8 & 20.3 &
 1.279 & 1.273 &  44.0 \cr
110357.71+031808.2 & 110401.49+031817.5 & 18.2 & 19.1 &
 1.941 & 1.922 &  57.3 \cr
110838.31+255521.4 & 110838.99+255613.2 & 18.0 & 17.7 &
 0.719 & 0.732 &  52.6 \cr
111610.68+411814.4 & 111611.73+411821.5 & 19.0 & 18.0 &
 3.002 & 2.985 &  13.8$^b$ \cr
113457.73+084935.2 & 113459.37+084923.2 & 18.9 & 19.2 &
 1.533 & 1.525 &  27.1 \cr
121840.48+501543.4 & 121841.00+501535.8 & 18.3 & 16.8 &
 1.458 & 1.455 &   9.1 \cr
122708.94+195751.8 & 122712.13+195804.2 & 18.2 & 18.0 &
 1.966 & 1.962 &  46.7 \cr
124819.28+370626.3 & 124823.35+370628.9 & 19.0 & 19.1 &
 1.525 & 1.518 &  48.8 \cr
140012.77+313454.1 & 140012.85+313452.7 & 20.0 & 19.6 &
 3.317 & 3.316 &   1.7$^c$ \cr
152107.91+562448.4 & 152110.12+562436.0 & 19.1 & 19.8 &
 1.578 & 1.566 &  22.2 \cr
155217.93+045646.7 & 155218.08+045635.2 & 18.7 & 18.4 &
 1.563 & 1.564 &  11.7 \cr
165501.31+260517.4 & 165502.01+260516.5 & 17.9 & 17.8 &
 1.880 & 1.891 &   9.6 \cr
215727.26+001558.4 & 215728.35+001545.5 & 20.5 & 19.2 &
 2.570 & 2.553 &  20.8 \cr
\enddata

\tablenotetext{a}{The quasar pairs were selected by a redshift difference of
less than~0.02 and an angular separation ($\Delta \theta$)
less than 60$''$.}

\tablenotetext{b}{See Hennawi et al. (2010).}

\tablenotetext{c}{Known gravitational lens (Inada et al. 2008).}

\end{deluxetable}

\clearpage
%
%\begin{deluxetable}{rrrrrrrr}
%%\tabletypesize{\small}
%\tablewidth{0pt}
%\tablecaption {Modal Quasar Colors as a Function of Redshift$^{a}$}
%\tablehead{
%\colhead{$z_{\rm bin}^{b}$} &
%\colhead{$\langle z \rangle^{c} $} &
%\colhead{$N_{\rm QSO}$} &
%\colhead{$(g-i)$} &
%\colhead{$(u-g)$} &
%\colhead{$(g-r)$} &
%\colhead{$(r-i)$} &
%\colhead{$(i-z)$}
%}
%
%\startdata
%  0.12 &0.137 & 238 & 0.498 &$-$0.061 & 0.097 & 0.435 & $-$0.095 \\
%  0.18 &0.181 & 242 & 0.560 &$-$0.076 & 0.169 & 0.374 & $-$0.067 \\
%  0.21 &0.211 & 375 & 0.579 & 0.018 & 0.223 & 0.356 & $-$0.043 \\
%  0.23 &0.230 & 165 & 0.524 &$-$0.032 & 0.142 & 0.331 & $-$0.020 \\
%  0.24 &0.240 & 189 & 0.580 & 0.021 & 0.271 & 0.281 & 0.086 \\
%\enddata
%
%\tablenotetext{a}{
%Table 5 is presented in its entirety in the electronic edition of the
%Astronomical Journal.  A portion is shown here for guidance regarding
%its form and content.}
%\tablenotetext{b}{
%The value of the redshift in the middle of the bin.}
%\tablenotetext{c}{
%The value of the average redshift of the quasars in the bin.}
%
%\end{deluxetable}
%
%\clearpage

\begin{deluxetable}{crrrcrrrrrrrrrr}
\tabletypesize{\small}
\rotate
\tablewidth{0pt}
\tablecaption {Quasars Lacking CAS Photometry$^{a}$}
\tablehead{
\colhead{Object (SDSS J)} &
\colhead{R.A. (deg)} &
\colhead{Dec (deg)} &
\colhead{Redshift} &
\multicolumn{2}{c}{$u$} &
\multicolumn{2}{c}{$g$} &
\multicolumn{2}{c}{$r$} &
\multicolumn{2}{c}{$i$} &
\multicolumn{2}{c}{$z$} &
\colhead{$A_u$}
}

\startdata
000439.57$-$004319.2 &  1.164883 & $-$0.722014 & 2.4211 & 21.475 & 0.179 &
 20.633 & 0.039 & 20.018 & 0.028 & 19.751 & 0.031 & 19.484 & 0.076  & 0.076 \\
001509.46+351111.6 &  3.789439 & 35.186576 & 2.4736 & 20.177 & 0.040 &
 19.539 & 0.019 & 19.319 & 0.019 & 19.218 & 0.019 & 18.960 & 0.041  & 0.041 \\
001510.37+354004.9 &  3.793231 & 35.668044 & 2.9886 & 20.980 & 0.087 &
 19.467 & 0.023 & 19.187 & 0.024 & 19.153 & 0.026 & 19.002 & 0.042  & 0.042 \\
001552.04+352216.4 &  3.966839 & 35.371239 & 0.7361 & 19.225 & 0.025 &
 19.003 & 0.016 & 19.037 & 0.014 & 19.082 & 0.018 & 18.956 & 0.039  & 0.039 \\
001628.24+344434.9 &  4.117698 & 34.743040 & 1.2350 & 18.938 & 0.024 &
 18.909 & 0.014 & 18.709 & 0.013 & 18.709 & 0.014 & 18.803 & 0.043  & 0.043 \\
\enddata

\tablenotetext{a}{
Table 5 is presented in its entirety in the electronic edition of the
Astronomical Journal.  A portion is shown here for guidance regarding
its form and content.}

\end{deluxetable}

\clearpage

\end{document}